\documentclass[10pt,journal]{IEEEtran}

\usepackage{amssymb}
\usepackage{amsmath}
\usepackage{cite}
\usepackage{url}
\usepackage{xcolor}
\usepackage{cite,graphicx,amsmath,amssymb}
\usepackage{fancyhdr}
\usepackage{mdwmath}
\usepackage{mdwtab}
\usepackage{caption}
\usepackage{amsthm}
\usepackage{setspace}
\usepackage{hyperref}
\usepackage{algorithm}
\usepackage{algorithmic}
\usepackage{multirow}
\usepackage{makecell}
\usepackage{mathtools}
\usepackage{subcaption}
\usepackage{bm}
\usepackage{tikz}
\usepackage{xurl}
\usepackage{stfloats}
\usepackage{mathrsfs}

\usepackage{booktabs}
\usepackage{multirow}
\usepackage{siunitx}
\usepackage{balance}
\usepackage{etoolbox}

\hypersetup{colorlinks=true,
linkcolor=blue,
citecolor=blue,      
urlcolor=black,
}

\graphicspath{{./src/img/}}

\newtheorem{remark}{Remark}
\newtheorem{theorem}{Theorem}

\newtheorem{lemma}{Lemma}

\newtheorem{corollary}{Corollary}

\newtheorem{proposition}{Proposition}

\allowdisplaybreaks
\NewDocumentCommand{\multiubrace}{mmm}
 {
  \egreg_multiubrace:nnn {#1} {#2} {#3}
 }

\title{Mutual Coupling in Continuous Aperture Arrays: Physical Modeling and Beamforming Design}

\author{
        Zhaolin Wang,~\IEEEmembership{Member, IEEE}, Kuranage Roche Rayan Ranasinghe,~\IEEEmembership{Graduate Student Member, IEEE},\\ Giuseppe Thadeu Freitas de Abreu,~\IEEEmembership{Senior Member, IEEE}, and Yuanwei Liu,~\IEEEmembership{Fellow, IEEE}
\thanks{An earlier version of this paper was presented in part at the IEEE International Conference on Communications (ICC), Glasgow, United Kingdom, 2026 \cite{conference_version}.}
\thanks{Z. Wang and Y. Liu are with the Department of Electrical and Computer Engineering, The University of Hong Kong, Hong Kong (e-mail: \{zhaolin.wang, yuanwei\}@hku.hk).}
\thanks{K. R. R. Ranasinghe and G. T. F. de Abreu are with
the School of Computer Science and Engineering, Constructor University
(previously Jacobs University Bremen), Campus Ring 1, 28759 Bremen,
Germany (e-mail: \{kranasinghe, gabreu\}@constructor.university).}
\vspace{-0.7cm}
}

\begin{document}

\maketitle
\begin{abstract}
    The phenomenon of mutual coupling in continuous aperture arrays (CAPAs) is studied. First, a general physical model for the phenomenon that accounts for both polarization and surface dissipation losses is developed. Then, the uni-polarized coupling kernel is characterized, revealing that polarization induces anisotropic coupling and invalidates the conventional half-wavelength spacing rule for coupling elimination. Next, the beamforming design problem for CAPAs with coupling is formulated as a functional optimization problem, leading to the derivation of optimal beamforming structures via the calculus of variations. To address the challenge of inverting the coupling kernel in the optimal structure, two methods are proposed: 1) the \emph{kernel approximation} method, which yields a closed-form solution via wavenumber-domain transformation and Gauss-Legendre quadrature, and 2) the \emph{conjugate gradient} method, which addresses an equivalent quadratic functional optimization problem iteratively. Furthermore, the optimal array gain and beampattern are analyzed at the large-aperture limit. Finally, the proposed continuous mutual coupling model is extended to spatially discrete arrays (SPDAs), and comprehensive numerical results are provided, demonstrating that: 1) coupled SPDA performance correctly converges to the CAPA limit, while uncoupled models are shown to violate physics, 2) polarization results in anisotropic array gain behavior, and 3) the coupled beampattern exhibits higher directivity than the uncoupled beampattern.
\end{abstract}


\section{Introduction} \label{sec:intro}

\IEEEPARstart{T}HE pursuit of next-generation wireless systems has driven significant research into antenna systems capable of unprecedented spatial resolution and connection density. A key paradigm in this evolution is the shift toward massive or even gigantic multiple-input multiple-output (MIMO) that exploits extremely large-scale or ultra dense antenna arrays \cite{bjornson2024enabling, 10220205, wang2024tutorial}, which fundamentally challenge traditional design principles. In this context, continuous aperture arrays (CAPAs) have emerged as a foundational concept \cite{liu2024capa}, representing the theoretical limit of a spatially discrete array (SPDA) as the antenna spacing shrinks to zero. Unlike SPDAs, CAPAs provide the potential to maximize aperture efficiency and form a basis for understanding the ultimate physical limits relying on the electromagnetic information theory \cite{liu2024capa, zhu2024electromagnetic, bjornson2024towards}.

A primary physical challenge in realizing the performance of any dense array, and especially a CAPA, is mutual coupling due to the mutual radiation of antennas. Mutual coupling has been a long-standing research topic in SPDAs \cite{1310320, 4201034, 5446312, masouros2013large, 10158708, yuan2023effects, 10309946, 10547020}, which is typically studied relying on the circuit and multiport network models.  Different assumptions are commonly made in these studies, such as specific antenna types, the absence of polarization effects, or lossless radiation surfaces, which can limit the general applicability of their findings.

In contrast to SPDAs, the mutual coupling effect in CAPAs has received significantly less attention. In particular, early work examined the spatial degrees of freedom between continuous electromagnetic (EM) volumes and surfaces via eigenfunction analysis \cite{miller2000communicating, dardari2020communicating, 9896943} and investigated corresponding capacity limits using methods such as Kolmogorov information theory and Fredholm determinant analysis \cite{jensen2008capacity, 8585146, wan2023mutual}. More recently, several studies have addressed beamforming for CAPAs. Owing to the continuous nature of the aperture, CAPA beamforming is formulated with Hilbert-Schmidt operators and functional optimization, which limits the direct reuse of designs developed for SPDAs. To address these challenges, several approaches have been proposed, including wavenumber-domain discretization based on Fourier plane-wave expansion \cite{9906802, zhang2023pattern}, calculus of variations leading to Fredholm integral equations \cite{10910020, 11122426}, and subspace expansion \cite{guo2024deep}. However, mutual coupling is not effectively addressed in the aforementioned works. While a recent study provided a wavenumber-domain analysis of CAPA mutual coupling \cite{11006094}, it relied on simplifications of omitting both polarization effects and surface dissipation loss, and it did not provide an explicit beamforming solution. Consequently, a complete CAPA beamforming design that explicitly and rigorously accounts for mutual coupling is still lacking.

Against the above background, this paper tackles the intricate problem of mutual coupling in CAPA beamforming, with the consideration of both polarization and surface dissipation loss. In particular, a fundamental challenge in this problem is characterizing the inverse of the mutual coupling kernel \cite{11006094, 11122426}, which lacks an explicit solution. This paper effectively addresses this challenge by proposing a \emph{kernel approximation} method and a \emph{conjugate gradient} method. The main contributions are summarized as follows:
\begin{itemize}
    \item A general physical model for mutual coupling in CAPAs is developed, which accounts for both polarization and surface dissipation loss. Based on this model, we in particular characterize the uni-polarized mutual coupling kernel, revealing that polarization leads to anisotropic mutual coupling. This finding demonstrates that the conventional half-wavelength spacing rule for eliminating mutual coupling is no longer applicable.
    \item Two design methods for CAPA beamforming under mutual coupling are proposed, namely the \emph{kernel approximation} and the \emph{conjugate gradient} methods, both of which address the challenges of inverting the coupling kernel. In particular, the kernel approximation method approximates the true kernel in the wavenumber domain using Gauss-Legendre quadrature, leading to a closed-form solution for the inverse. In contrast, the conjugate gradient method obtains the beamforming solution iteratively by solving an equivalent quadratic functional optimization problem.
    \item An analysis of both the array gain and the beampattern of CAPA beamforming is offered, yielding the closed-form array gain in the large-aperture limit and revealing the necessity of accounting for mutual coupling. A comparison of the beampatterns of coupled and uncoupled CAPAs further reveals that, under a large-aperture approximation, the coupled beampattern is essentially filtered by the wavenumber-domain mutual coupling kernel relative to the uncoupled case.
    \item An extension of the proposed continuous mutual coupling model to SPDAs is given, including the derivation of the discrete mutual coupling matrix as a function of both the continuous coupling kernel and the current profile of each antenna element, from which the optimal discrete beamformer is obtained.
    \item Comprehensive numerical results to evaluate the CAPA beamforming performance are shown, which unveils the following insights: 1) When mutual coupling is considered, the array gain of an SPDA converges to the CAPA limit as the antenna spacing shrinks. In contrast, uncoupled SPDA models show gains growing unboundedly, leading to a violation of physics. 2) Polarization leads to different beamforming behavior in different directions. In particular, for directions in the plane aligned with the polarization, the array gain is maximized at front-fire and vanishes at end-fire. For the direction in the orthogonal plane, the array gain exhibits peaks at both front-fire and end-fire, though the end-fire peaks diminish as the aperture size increases. 3) Compared to the uncoupled beampattern, the coupled beampattern exhibits higher directivity for both front-fire and end-fire beamforming.
\end{itemize}

The remainder of this paper is organized as follows. Section II develops the system model and derives both the tri-polarized and uni-polarized coupling kernels. Section III formulates the beamforming design problem and presents the kernel approximation and conjugate gradient methods. Section IV analyzes the resulting array gain and beampatterns. Section V extends the proposed continuous model to the SPDA case. Section VI provides numerical results to validate our analysis and methods, and Section VII concludes the paper.

\emph{Notations:} Scalars, vectors/matrices, and Euclidean subspaces are denoted by regular, boldface, and calligraphic letters, respectively. The sets of complex, real, and integer numbers are represented by $\mathbb{C}$, $\mathbb{R}$, and $\mathbb{Z}$, respectively. The inverse, transpose, conjugate transpose, and trace operations are represented by $(\cdot)^{-1}$, $(\cdot)^T$, $(\cdot)^H$, and $\mathrm{Tr}(\cdot)$, respectively. The absolute value and Euclidean norm are indicated by $|\cdot|$ and $\|\cdot\|$, respectively. The Lebesgue measure of a Euclidean subspace $\mathcal{S}$ is denoted by $|\mathcal{S}|$. The real part of a complex number is denoted by $\Re \{\cdot\}$. An identity matrix of size $N \times N$ is denoted by $\mathbf{I}_N$. The Dirac delta function on the space $\mathbb{R}^{N \times 1}$ is denoted by 
\begin{equation}
    \begin{aligned} \label{identify_function}
        \delta(\mathbf{s} - \mathbf{z}) &= 0 \quad \text{ for } \mathbf{s} \neq \mathbf{z}, \\
        \int_{\mathcal{V}} \delta(\mathbf{s} - \mathbf{z}) d \mathbf{s} &= 1,
    \end{aligned}
\end{equation}
where $\mathbf{s} \in \mathbb{R}^{N \times 1}$ and $\mathbf{z} \in \mathbb{R}^{N \times 1}$, and $\mathcal{V} \subseteq \mathbb{R}^{N \times 1}$ is any volume that contains the point $\mathbf{s} = \mathbf{z}$.

\vspace{-0.1cm}
\section{System Model} \label{sec:model}

\vspace{-0.1cm}
\subsection{Mutual Coupling}

Mutual coupling arises because different current distributions interact through the EM fields they generate, thereby affecting the required driving power and the resulting radiation behavior. To characterize mutual coupling in CAPAs, let us consider a transmit surface $\mathcal{S}$, which contains a time-harmonic source current to radiate EM waves into free space. In particular, we consider a rectangular planar transmit surface, which is specified by 
\begin{equation} \label{transmit_aperture}
    \mathcal{S} = \left\{ [s_x, s_y, 0]^T\, \left|\, |s_x| \le \frac{L_x}{2}, |s_y| \le \frac{L_y}{2} \right. \right\}.
\end{equation}

Let $\bm{j}_{\mathrm{t}}(\mathbf{s}) \in \mathbb{C}^{3 \times 1},\mathbf{s} \in \mathcal{S}$, denote the Fourier transform of the source current density at frequency $f$, where the corresponding wavelength is $\lambda$. The time factor is assumed to be $e^{-j 2 \pi f t}$.  We focus on a narrowband single-carrier system. Wideband multi-carrier systems can be handled by applying the proposed model to each carrier frequency independently.

The total EM power, denoted by $P_{\mathrm{em}}$, exerted by the source current is divided into two fundamental components: the power radiated into free space and the power dissipated as heat due to the finite conductivity of the surface, as discussed below.

\subsubsection{Radiated Power}
The free-space radiated field caused by the source current $\bm{j}_{\mathrm{t}}(\mathbf{s})$ can be characterized by solving the inhomogeneous Helmholtz wave equation, leading to the solution given by \cite{dardari2020communicating, bjornson2024towards, 11006094}
\begin{equation} \label{radiation_field}
    \bm{e}_{\mathrm{rad}}(\mathbf{s}) = \int_{\mathcal{S}} \mathbf{G}(\mathbf{s} - \mathbf{z}) \bm{j}_{\mathrm{t}}(\mathbf{z}) d \mathbf{z},
\end{equation}
where $\mathbf{G}(\mathbf{s}) \in \mathbb{C}^{3 \times 3}$ is the dyadic Green's function:
\begin{equation} \label{Green_function}
    \mathbf{G}(\mathbf{s}) = -j \kappa_0 Z_0 \left( \mathbf{I}_3 + \frac{1}{\kappa_0^2} \nabla \nabla \right) g(\mathbf{s}).
\end{equation} 
Here, $\kappa_0 = 2\pi/\lambda$ denotes the wavenumber, $Z_0 \approx 120 \pi$ denotes the free space impedance, $\nabla$ is nabla operator for $\mathbf{s}$, and $g(\mathbf{s})$ is the scalar Green's function given by 
\begin{equation}
    g(\mathbf{s}) = \frac{e^{j \kappa_0 \|\mathbf{s}\|}}{4 \pi \|\mathbf{s}\|}.
\end{equation}

The radiated power is the time-averaged work done by the source current working against this radiated field across the transmit surface \cite{11006094}:
\begin{align} \label{tri_radiation_power}
    P_{\mathrm{rad}} & =  \frac{1}{2} \Re \left\{ \int_{\mathcal{S}} \bm{j}^H_{\mathrm{t}}(\mathbf{s}) \bm{e}_{\mathrm{rad}} (\mathbf{s}) d \mathbf{s} \right\} \nonumber \\
    & = \frac{1}{2} \Re \left\{ \int_{\mathcal{S}} \int_{\mathcal{S}}  \bm{j}_{\mathrm{t}}^H(\mathbf{s}) \mathbf{G}(\mathbf{s} - \mathbf{z}) \bm{j}_{\mathrm{t}}(\mathbf{z}) d \mathbf{z} d \mathbf{s}  \right\} \nonumber \\
    & = \frac{1}{2} \int_{\mathcal{S}} \int_{\mathcal{S}}  \bm{j}_{\mathrm{t}}^H(\mathbf{s}) \Re \left\{\mathbf{G}(\mathbf{s} - \mathbf{z})\right\} \bm{j}_{\mathrm{t}}(\mathbf{z}) d \mathbf{z} d \mathbf{s}, 
\end{align}
where the last step is obtained following \cite[Appendix B]{11006094}. The real part $\Re\{\cdot\}$ appears because the imaginary part corresponds to reactive, non-radiative field exchange, and therefore does not contribute to the time-averaged radiated power.

In existing literature \cite{5446312, 9906802, zhang2023pattern}, the radiated power of an EM wave is typically characterized from the perspective of the Poynting vector. As shown in Appendix \ref{Poynting_vector_proof}, this method essentially yields the same result shown in \eqref{tri_radiation_power}. 

\subsubsection{Surface Dissipation Loss}
Practical transmit surfaces are not perfect conductors. They exhibit an inherent surface resistance $Z_{s}\in\mathbb{R}$. Under the assumption that the transmit surface is a good conductor, the surface resistance can be expressed as \cite[Eq. (1.125)]{pozar2021microwave}
\begin{equation}
    Z_s = \sqrt{\frac{\pi f \mu_s}{\sigma_s}},
\end{equation}
where $\mu_s$ and $\sigma_s$ are the surface permeability and conductivity, respectively. 

Due to this resistance, a portion of the supplied energy is lost strictly to local Joule heating, leading to the following dissipated power \cite[Eq. (1.131)]{pozar2021microwave}:
\begin{equation}
    P_{\mathrm{diss}} = \frac{Z_s}{2}  \int_{\mathcal{S}} \|\bm{j}_{\mathrm{t}}(\mathbf{s})\|^{2} d \mathbf{s}.
\end{equation}

\subsubsection{Coupling Kernel}
The total EM power is the sum of both components, calculated by $P_{\mathrm{em}} = P_{\mathrm{rad}} + P_{\mathrm{diss}}$. By factoring out the source current density, we can define a unified coupling kernel $\mathbf{C}(\mathbf{s})\in\mathbb{C}^{3\times3}$ as
\begin{equation}
    \mathbf{C}(\mathbf{s}) = Z_s \delta(\mathbf{s}) \mathbf{I}_3 + \mathbf{G}(\mathbf{s}),
\end{equation} 
where $\delta(\mathbf{s})$ is the three-dimensional (3D) Dirac delta function. More particularly, the first term, $Z_s \delta(\mathbf{s}) \mathbf{I}_3$, is a purely local contribution, which models ohmic dissipation at each surface point. In contrast, $\mathbf{G}(\mathbf{s})$ is nonlocal, which captures how currents at one location induce electric fields that affect the power required to sustain currents at other locations.

Utilizing this unified coupling kernel, the total EM power can be expressed in a single integral equation as follows:
\begin{align} 
    P_{\mathrm{em}} = \frac{1}{2} \int_{\mathcal{S}} \int_{\mathcal{S}}  \bm{j}_{\mathrm{t}}^H(\mathbf{s}) \Re \left\{\mathbf{C}(\mathbf{s} - \mathbf{z})\right\} \bm{j}_{\mathrm{t}}(\mathbf{z}) d \mathbf{z} d \mathbf{s}.  
\end{align}

\begin{remark}
    \normalfont
    \emph{(Types of Coupling)}
    As shown in \eqref{tri_radiation_power}, the mutual coupling effect is characterized by the real part of the Green's function $\mathbf{G}(\mathbf{s})$, which is a full matrix as detailed in \eqref{Green_function}. This implies two distinct types of mutual coupling, namely \emph{inter-position} coupling arising from its spatial dependencies and \emph{inter-polarization} coupling arising from its non-diagonal matrix structure.
\end{remark}

\begin{remark} \label{remark_uncoupled}
    \normalfont
    \emph{(Uncoupled CAPA)} A CAPA is defined as uncoupled if its coupling kernel is a scaled Dirac delta function, i.e., $\mathbf{C}(\mathbf{s}) = \rho \delta(\mathbf{s}) \mathbf{I}_3$, where $\rho$ is a normalization factor. The transmit EM power then simplifies to $P_{\mathrm{em}} = \frac{\rho }{2} \int_{\mathcal{S}} \|\boldsymbol{j}_{\mathrm{t}}(\mathbf{s})\|^2 d \mathbf{s}$ by using the property of $\delta(\mathbf{s})$. This simplified power form, while effective for bounding transmit power \cite{9906802, zhang2023pattern} and simplifying waveform design \cite{10910020, 10938678} is, however, only a mathematical simplification rather than a model derived from physical principles, which can therefore be inaccurate.      
\end{remark}


\subsection{Uni-Polarized Coupling Kernel}

In the sequel, we focus on the inter-position coupling and specialize the general model to a fixed uni-polarized model in order to obtain explicit analytical results. Without loss of generality, the polarization direction is assumed to be aligned with the $y$-axis. The same projection-based method can be applied to any other fixed polarization direction, whereas a genuinely multi-polarized design would require retaining the full matrix-valued coupling model and is beyond the scope of this paper. Under this assumption, the source current is simplified into
\begin{equation}
    \bm{j}_{\mathrm{t}}(\mathbf{s}) = w(\mathbf{s}) \mathbf{u}_y,
\end{equation}
where $w(\mathbf{s}) \in \mathbb{C}$ is the uni-polarized component and $\mathbf{u}_y = [0,1,0]^T$ is the unit vector along with $y$-axis. 

For the uni-polarized source current, the transmit EM power becomes
\begin{align} \label{em_power}
    P_{\mathrm{em}} & =  \frac{1}{2} \int_{\mathcal{S}} \int_{\mathcal{S}} w^*(\mathbf{s}) \Re \left\{ \mathbf{u}_y^T \mathbf{C}(\mathbf{s} - \mathbf{z}) \mathbf{u}_y \right\} w(\mathbf{z}) d \mathbf{z} d \mathbf{s}  \nonumber \\ 
    & =  \frac{1}{2} \int_{\mathcal{S}} \int_{\mathcal{S}} w^*(\mathbf{s}) c(\mathbf{s} - \mathbf{z}) w(\mathbf{z}) d \mathbf{z} d \mathbf{s}.
\end{align} 
Here, $c(\mathbf{s}) \in \mathbb{R}$ is the scalar coupling kernel, given by 
\begin{align} \label{scalar_coupling}
    c(\mathbf{s}) & = \Re \left\{ \mathbf{u}_y^T \mathbf{C}(\mathbf{s}) \mathbf{u}_y \right\} \nonumber \\
    & = \underbrace{ \vphantom{\frac{1}{\kappa_0^2}} Z_s \delta(\mathbf{s})}_{\text{dissipation}} + \underbrace{\kappa_0 Z_0 \left( \varphi(\mathbf{s}) + \frac{1}{\kappa_0^2} \partial_y^2 \varphi(\mathbf{s}) \right)}_{\text{radiation}}, 
\end{align}
where
\begin{equation}
    \varphi(\mathbf{s}) = \frac{\sin(\kappa_0 \|\mathbf{s}\|)}{4\pi \|\mathbf{s}\|}.
\end{equation}
For notational convenience, we define the radiation mutual coupling kernel as 
\begin{equation} \label{radiation_mutual_coupling_kernel}
    c_{\mathrm{rad}} (\mathbf{s}) = \kappa_0 Z_0 \left( \varphi(\mathbf{s}) + \frac{1}{\kappa_0^2} \partial_y^2 \varphi(\mathbf{s}) \right).
\end{equation}

\begin{remark}
    \normalfont

    \emph{(Impact of Polarization)} The radiation mutual coupling kernel $c_{\mathrm{rad}} (\mathbf{s})$ comprises two components: the function $\varphi(\mathbf{s})$ and its second-order derivative $\partial_y^2 \varphi(\mathbf{s})$ along the polarization direction. This finding is partially consistent with existing studies on the mutual coupling of idealized isotropic antennas \cite{11006094, 5446312, yordanov2009arrays, 6404701, friedlander2020extended}, which characterize coupling using only the $\varphi(\mathbf{s})$ term. However, purely isotropic antennas are not physically realizable, making polarization a necessary consideration for practical systems. The result in \eqref{radiation_mutual_coupling_kernel} reveals that the impact of polarization is characterized by the second-order derivative $\partial_y^2 \varphi(\mathbf{s})$ in the polarization direction.
\end{remark}

\begin{figure}[t!]
  \centering
  \includegraphics[width=0.45\textwidth]{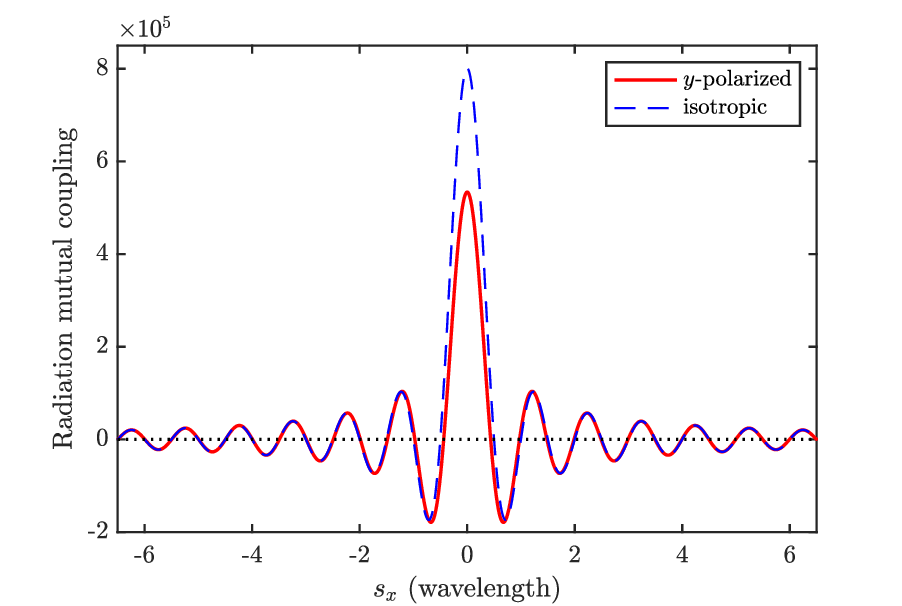}
  \caption{Radiation mutual coupling kernel $c_{\mathrm{rad}}(\mathbf{s})$ along the $x$-axis ($s_y = 0$) at $2.4$ GHz, comparing the proposed polarized model in \eqref{radiation_mutual_coupling_kernel} with the isotropic-kernel model that omits the $\partial_y^2 \varphi(\mathbf{s})$ term. The null locations are only slightly shifted from the conventional half-wavelength rule.}
  \label{fig_mutual_coupling_x_axis}

  \centering
  \includegraphics[width=0.45\textwidth]{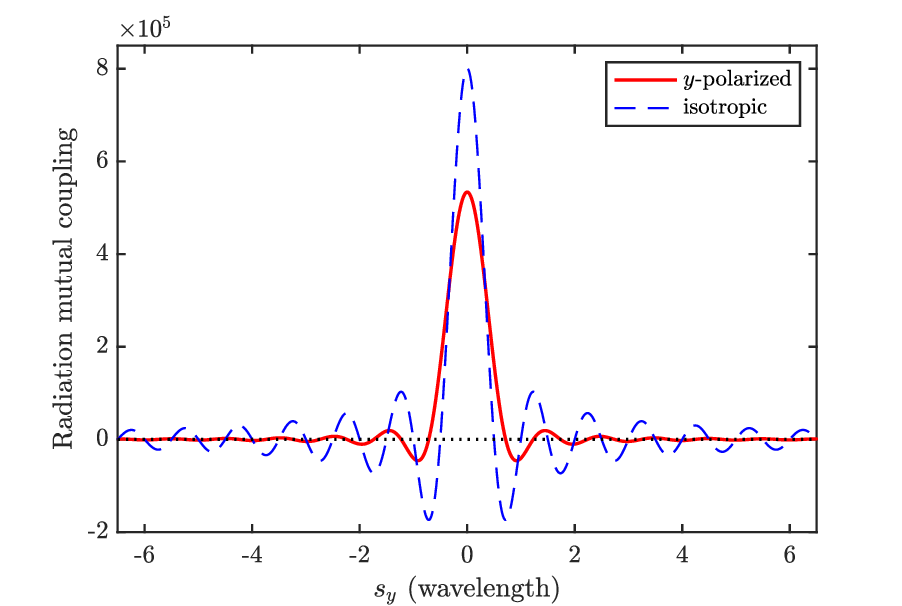}
  \caption{Radiation mutual coupling kernel $c_{\mathrm{rad}}(\mathbf{s})$ along the $y$-axis ($s_x = 0$) at $2.4$ GHz, comparing the proposed polarized model in \eqref{radiation_mutual_coupling_kernel} with the isotropic-kernel model that omits the $\partial_y^2 \varphi(\mathbf{s})$ term. In the polarization direction, the null locations are significantly shifted and the sidelobes are substantially reduced.}

  \label{fig_mutual_coupling_y_axis}
\end{figure}

Fig. \ref{fig_mutual_coupling_x_axis} and Fig. \ref{fig_mutual_coupling_y_axis} illustrate the impact of polarization on mutual coupling at $2.4$ GHz by comparing the full model against the simplified isotropic antenna model, which omits the second-order derivative $\partial_y^2 \varphi(\mathbf{s})$.
 A key conclusion for isotropic antennas is that mutual coupling vanishes when $\varphi(\mathbf{s}) = 0$, which occurs at antenna spacings of integer multiples of a half-wavelength, i.e., $\|\mathbf{s}\| = i \lambda/2, i \in \mathbb{Z}$. 
%

However, incorporating polarization invalidates this conclusion. Specifically, Fig. \ref{fig_mutual_coupling_x_axis} shows that along the $x$-axis (i.e., orthogonal to the polarization), the polarization primarily affects the mainlobe, while the coupling nulls are only slightly shifted from the isotropic case. In contrast, Fig. \ref{fig_mutual_coupling_y_axis} shows that along the $y$-axis (i.e., the polarization direction), the nulls are substantially shifted, and the sidelobe levels are significantly reduced. The following proposition provides further details on the nulls of the full mutual coupling kernel.
\begin{proposition} \label{proposition_null}
    \normalfont
    In polar coordinates $(r, \psi)$, defined by $s_x = r \sin \psi$ and $s_y = r \cos \psi$, the nulls of the radiation mutual coupling kernel, i.e., $c_{\mathrm{rad}}(\mathbf{s}) = 0$, are found by solving the following transcendental equation for $\epsilon = \kappa_0 r$, given by
    \begin{equation} \label{null_condition}
        \sin^2 \psi \left( (\epsilon^2 - 3 ) \sin\epsilon + 3 \epsilon \cos \epsilon \right) + 2 (\sin \epsilon - \epsilon  \cos \epsilon) = 0.
    \end{equation}
\end{proposition}
\begin{IEEEproof}
    Please refer to Appendix \ref{proposition_null_proof}.
\end{IEEEproof}
Based on \textbf{Proposition \ref{proposition_null}}, we can analyze the nulls of $c_{\mathrm{rad}}(\mathbf{s})$ along the primary axes. The kernel nulls along the $x$-axis are found by setting $\psi = \pi/2$ (i.e., $s_y = 0$) in \eqref{null_condition}, which yields
\begin{equation}
    (\epsilon^2 - 1) \sin \epsilon + \epsilon \cos \epsilon = 0.
\end{equation}     
The first few positive roots are $\epsilon \approx 2.74, 6.12,\, 9.32,\, \dots$, which correspond to normalized spacings of
\begin{equation}
    \frac{|s_x|}{\lambda} = \frac{\epsilon}{2\pi} \approx 0.44,\, 0.97,\, 1.48, \dots
\end{equation} 
These results confirm the simulation in Fig. \ref{fig_mutual_coupling_x_axis}, showing that the nulls are only slightly shifted from the integer half-wavelengths (i.e., $0.5, 1.0, 1.5, \dots$) of the isotropic case.
Similarly, the nulls along the $y$-axis are found by setting $\psi = 0$ (i.e., $s_x = 0$) in \eqref{null_condition}, yielding
\begin{equation}
    \tan \epsilon - \epsilon = 0.
\end{equation}
The first few positive roots are $\epsilon \approx 4.50,\, 7.73,\, 10.90, \dots$, corresponding to normalized spacings of
\begin{equation}
    \frac{|s_y|}{\lambda} = \frac{\epsilon}{2\pi} \approx 0.72,\, 1.23,\, 1.73, \dots
\end{equation}
These values validate the observation from Fig. \ref{fig_mutual_coupling_y_axis}, confirming that the nulls in the polarization direction are significantly shifted from the isotropic case.

\subsection{Channel Model}

Let $\mathbf{r} \in \mathbb{R}^{3 \times 1}$ and $\mathbf{u}_{\mathrm{r}} \in \mathbb{R}^{3 \times 1}$ denote the location and polarization direction of the receiver, respectively. To maximize the signal reception performance, we assume that the polarization is perfectly matched, i.e., $\mathbf{u}_{\mathrm{r}} = \mathbf{u}_y$. The effective electric field captured by the receiver is given by
\begin{align} \label{receive_electric_field}
    e_{\mathrm{r}} = & \mathbf{u}_y^T \bm{e}_{\mathrm{rad}}(\mathbf{r}) = \int_{\mathcal{S}} h(\mathbf{s}) w(\mathbf{s}) d \mathbf{s}.
\end{align}   
Here, $h(\mathbf{s}) \in \mathbb{C}$ is the uni-polarized channel response, given by 
\begin{align} \label{near_field_channel}
    h(\mathbf{s}) & = \mathbf{u}_y^T \mathbf{G}(\mathbf{r} - \mathbf{s}) \mathbf{u}_y \nonumber \\
    & = -j \kappa_0 Z_0 \left( g(\mathbf{r} - \mathbf{s}) + \frac{1}{\kappa_0^2} \partial_y^2 g(\mathbf{r} - \mathbf{s}) \right).
\end{align} 

This work focuses on a receiver located in the far-field. For a general far-field direction, define the direction vector $\mathbf{d}(\theta, \phi) = [\cos \theta \sin\phi, \sin \theta \sin \phi, \cos \phi]^T$, the associated 3D wavevector $\mathbf{k} = \kappa_0 \mathbf{d}(\theta, \phi) = [\kappa_x, \kappa_y, \kappa_z]^T$, and its 2D projection onto the transmit-surface plane $\boldsymbol{\kappa} = [\kappa_x, \kappa_y, 0]^T$, with $\kappa_z(\boldsymbol{\kappa}) = \sqrt{\kappa_0^2 - \kappa_x^2 - \kappa_y^2}$. For the fixed receiver direction $(\theta_0, \phi_0)$, we further define $\mathbf{k}_{\mathrm{r}} \triangleq \kappa_0 \mathbf{d}(\theta_0, \phi_0) = [\kappa_{\mathrm{r},x}, \kappa_{\mathrm{r},y}, \kappa_{\mathrm{r},z}]$ and $\boldsymbol{\kappa}_{\mathrm{r}} \triangleq [\kappa_{\mathrm{r},x}, \kappa_{\mathrm{r},y}, 0]^T$. The receiver position is $\mathbf{r} = R_0 \mathbf{d}(\theta_0, \phi_0)$, where $R_0$ is the distance from the origin.
Under far-field assumptions, the spherical wave represented by the Green's function can be approximated as a plane wave across the transmit surface as~\cite{10220205}
\begin{equation} \label{far_field_scalar_Green}
    g(\mathbf{r} - \mathbf{s}) = \frac{e^{j \kappa_0 R_0}}{4 \pi R_0} e^{-j \mathbf{k}_{\mathrm{r}}^T \mathbf{s}} = \frac{e^{j \kappa_0 R_0}}{4 \pi R_0} e^{-j \boldsymbol{\kappa}_{\mathrm{r}}^T \mathbf{s}},
\end{equation}
where we have $\mathbf{k}_{\mathrm{r}}^T \mathbf{s} = \boldsymbol{\kappa}_{\mathrm{r}}^T \mathbf{s}$ since $\mathbf{s} = [s_x, s_y, 0]^T \in \mathcal{S}$.
As a consequence, the far-field approximation of the channel response $h(\mathbf{s})$ is given by 
\begin{equation} \label{far_field_channel}
    h(\mathbf{s}) = \beta_{\mathrm{r}} e^{-j\boldsymbol{\kappa}_{\mathrm{r}}^T \mathbf{s} },
\end{equation}  
where $\beta_{\mathrm{r}} \triangleq \beta(\theta_0,\phi_0, R_0)$ is a complex constant that incorporates the path loss and the effect of polarization projection, characterized by
\begin{equation} \label{beta_factor}
    \beta(\theta, \phi, R) = \frac{-j \kappa_0 Z_0 e^{j \kappa_0 R}}{4 \pi R} \left( 1 - \sin^2 \theta \sin^2 \phi \right).
\end{equation}

\section{CAPA Beamforming Design with \\ Mutual Coupling} \label{optimization_method}

\subsection{Optimal Beamforming}

The objective of beamforming design is to optimize $w(\mathbf{s})$ to maximize the signal strength at a target receiver, and consequently, the fundamental single-user capacity limit, subject to a constraint on the total transmit power $P_{\mathrm{t}}$. This can be formulated as the following functional optimization problem:
\begin{subequations} \label{problem_single_user}
    \begin{align}
        \max_{w(\mathbf{s})} \quad & \left| \int_{\mathcal{S}} h(\mathbf{s}) w(\mathbf{s}) d \mathbf{s} \right|^2 \\
        \label{power_constraint}
        \mathrm{s.t.} \quad & \frac{1}{2} \int_{\mathcal{S}} \int_{\mathcal{S}} w(\mathbf{s}) c(\mathbf{s} - \mathbf{z}) w^*(\mathbf{z}) d \mathbf{z} d \mathbf{s} \le P_{\mathrm{t}}.
    \end{align}
\end{subequations}
The following derivation is written for the scalar beamformer induced by the uni-polarized reduction in Section~\ref{sec:model}. Extending the design to a genuinely multi-polarized system would require a vector-valued functional optimization based on the full matrix-valued coupling kernel $\mathbf{C}(\mathbf{s})$.
The solution to this problem is characterized by the following theorem.

\begin{theorem} \label{theorem_1}
    \normalfont 
    The optimal solution to problem \eqref{problem_single_user} is given by
    \begin{equation} \label{optimal_solution_SU}
        w_{\mathrm{opt}}(\mathbf{s}) = \sqrt{\frac{2 P_{\mathrm{t}}}{\int_{\mathcal{S}} h(\mathbf{z}) v(\mathbf{z}) d \mathbf{z}}} v(\mathbf{s}),
    \end{equation}
    where $v(\mathbf{s})$ is the solution to the Fredholm integral equation 
    \begin{equation} \label{Fredholm_equation}
        \int_{\mathcal{S}} c(\mathbf{s} - \mathbf{z}) v(\mathbf{z}) d \mathbf{z} = h^*(\mathbf{s}), \, \forall \mathbf{s} \in \mathcal{S}.
    \end{equation}

\end{theorem}

\begin{IEEEproof}
    Please refer to the Appendix \ref{theorem_1_proof}
\end{IEEEproof}

\textbf{Theorem \ref{theorem_1}} reveals that the optimal beamformer is determined by the solution to the Fredholm integral equation in \eqref{Fredholm_equation}. In theory, this equation can be solved by convolving both sides with an inverse kernel $c(\mathbf{s})$ that satisfies
\begin{equation} \label{inverse_condition}
    \int_{\mathcal{S}} c^{-1}(\mathbf{z}' - \mathbf{s})  c(\mathbf{s} - \mathbf{z}) d \mathbf{s} = \delta(\mathbf{z}'-\mathbf{z}). 
\end{equation} 
Consequently,  multiplying both sides of \eqref{Fredholm_equation} by $c^{-1}(\mathbf{z}' - \mathbf{s})$ and integrating over $d \mathbf{z}'$ yields
\begin{equation} \label{optimal_inverse_kernel_solution}
    v(\mathbf{z}') = \int_{\mathcal{S}} c^{-1}(\mathbf{z}' - \mathbf{s}) h^*(\mathbf{s}) d \mathbf{s}.
\end{equation} 

However, a closed-form expression for $c^{-1}(\mathbf{s})$ is generally unavailable, particularly because the radiation mutual coupling component of the kernel is non-trivial. This makes the direct computation of \eqref{optimal_inverse_kernel_solution} intractable.
In the sequel, a pair of methods are proposed to address this challenge, namely the \emph{kernel approximation} method and the \emph{conjugate gradient} method.

\subsection{Kernel Approximation}

In this subsection, the proposed kernel approximation method is presented.

\subsubsection{Wavenumber-domain Approximation}
The kernel approximation method aims to find a tractable approximation of the coupling kernel whose inverse $c^{-1}(\mathbf{s})$ can be expressed in closed form. This is achieved using the two-dimensional (2D) Fourier transform, defined as
\begin{equation}
    \mathcal{F}\{f\}(\boldsymbol{\kappa}) = \iint_{-\infty}^{+\infty} f(\mathbf{s})e^{-j \boldsymbol{\kappa}^T \mathbf{s}} d s_x d s_y,
\end{equation}
where $\boldsymbol{\kappa} = [\kappa_x, \kappa_y, 0]^T$. Applying this transform to the radiation mutual coupling kernel $c_{\mathrm{rad}}(\mathbf{s})$ yields its wavenumber-domain representation $C_{\mathrm{rad}}(\boldsymbol{\kappa})$, which is given by
\begin{equation} \label{WD_coupling_kernel}
    C_{\mathrm{rad}} (\boldsymbol{\kappa}) = \begin{dcases}
            \frac{Z_0 ( 1 - \kappa_y^2/\kappa_0^2 )}{2 \sqrt{ 1 - \|\boldsymbol{\kappa}\|^2/\kappa_0^2} },  & \|\boldsymbol{\kappa}\| \le \kappa_0, \\
            0, & \|\boldsymbol{\kappa}\| > \kappa_0.
    \end{dcases}
\end{equation}
This expression is derived using the Weyl identity \cite{chew1999waves} and the property that the spatial derivative operator $\partial^2_y$ corresponds to multiplication by $-\kappa_y^2$ after the Fourier transform. The detailed derivation of \eqref{WD_coupling_kernel} is given in Appendix \ref{appendix_derivation_WD_coupling_kernel}. The spatial kernel $c_{\mathrm{rad}}(\mathbf{s})$ is then recovered via the inverse 2D Fourier transform
\begin{align} \label{inverse_Fourier}
    c_{\mathrm{rad}}(\mathbf{s}) & = \frac{1}{(2\pi)^2} \iint_{\|\boldsymbol{\kappa}\| \le \kappa_0} C_{\mathrm{rad}}(\boldsymbol{\kappa}) e^{j \boldsymbol{\kappa}^T \mathbf{s}} d \boldsymbol{\kappa} \nonumber \\
    & = \frac{1}{(2\pi)^2} \int_{-\kappa_0}^{\kappa_0} \int_{-\sqrt{\kappa_0^2 - \kappa_x^2}}^{+\sqrt{\kappa_0^2 - \kappa_x^2}} C_{\mathrm{rad}}(\boldsymbol{\kappa}) e^{j \boldsymbol{\kappa}^T \mathbf{s}} d \kappa_y d \kappa_x.
\end{align}

To obtain the closed-form inverse, we exploit the Gauss-Legendre quadrature to approximate the continuous integration in \eqref{inverse_Fourier} via a discrete summation. The Gauss-Legendre quadrature takes the form \cite{olver2010nist}
\begin{equation} \label{GL_approx}
    \int_{a}^{b} g(x) dx \approx \frac{b-a}{2} \sum_{m=1}^M \omega_m g \left(\frac{b-a}{2} \theta_m + \frac{a+b}{2} \right),
\end{equation}
where $M$ is the order of the Gauss-Legendre quadrature, and $\theta_m$ and $\omega_m$ denote the roots of the Gauss-Legendre polynomial and the corresponding weights, respectively. Given the geometric convergence of Gauss-Legendre quadrature for smooth integrands, this approximation is highly accurate with merely a few orders. 
Applying this quadrature to the inverse Fourier transform integral \eqref{inverse_Fourier} allows the radiation kernel $c_{\mathrm{rad}}(\mathbf{s})$ to be approximated as
\begin{align} \label{proposed_approx_kernel}
    c_{\mathrm{rad}}(\mathbf{s}) \approx \sum_{n=1}^{M} \sum_{m=1}^{M} \widetilde{\rho}_{nm} e^{j \widetilde{\boldsymbol{\kappa}}_{nm}^T \mathbf{s}},
\end{align}     
where
\begin{align}
    &\widetilde{\rho}_{nm} = \frac{W_n^{(x)} W_{nm}^{(y)}}{(2\pi)^2} C_{\mathrm{rad}} \left(\widetilde{\boldsymbol{\kappa}}_{nm} \right), \\
    &\widetilde{\boldsymbol{\kappa}}_{nm} = \left[ \kappa_{n}^{(x)}, \kappa_{nm}^{(y)}, 0 \right]^T.
\end{align}
More particularly, $(\kappa_{n}^{(x)}, W_n^{(x)})$ and $(\kappa_{nm}^{(y)}, W_{nm}^{(y)})$ are the scaled Gauss-Legendre coefficients for the integrals over $d\kappa_x$ and $d \kappa_y$, respectively, given by 
\begin{align}
    \kappa_{n}^{(x)} = \kappa_0 \theta_n, \quad \kappa_{nm}^{(y)} = \sqrt{\kappa_0^2- \left(\kappa_{n}^{(x)}\right)^2} \theta_m, \\
    W_n^{(x)} = \kappa_0 \omega_n, \quad W_{nm}^{(y)} = \sqrt{\kappa_0^2- \left(\kappa_{n}^{(x)}\right)^2} \omega_m.
\end{align}

By combining the approximation in \eqref{proposed_approx_kernel} with the dissipation term, the overall coupling kernel can be written in a separable form as
\begin{align} \label{approx_overall_kernel}
    c(\mathbf{s} - \mathbf{z}) & = Z_s \delta(\mathbf{s} - \mathbf{z}) + \sum_{n=1}^{M} \sum_{m=1}^{M} \widetilde{\rho}_{nm} e^{j \widetilde{\boldsymbol{\kappa}}_{nm}^T \mathbf{s}} e^{-j \widetilde{\boldsymbol{\kappa}}_{nm}^T \mathbf{z}} \nonumber \\
    & = Z_s \delta(\mathbf{s} - \mathbf{z}) + \sum_{i=1}^{J}  \rho_i e^{j \boldsymbol{\kappa}_i^T \mathbf{s}} e^{-j \boldsymbol{\kappa}_i^T \mathbf{z}}.
\end{align} 
In the final step, the double summation is re-indexed into a single sum of $J = M^2$ terms, where $\rho_i$ and $\boldsymbol{\kappa}_i$ are the re-indexed weights and wavenumber vectors, respectively. The inverse of the approximated coupling kernel is presented in the following proposition.

\begin{proposition} \label{proposition_0}
    \normalfont 

    The inverse of $c(\mathbf{s} - \mathbf{z})$ that satisfies the condition \eqref{inverse_condition} is given by 
    \begin{equation} \label{inverse_kernel}
        c^{-1}(\mathbf{z}' - \mathbf{s}) = \frac{1}{Z_s} \delta(\mathbf{z}' - \mathbf{s}) - \sum_{i=1}^J \sum_{l=1}^J \frac{\rho_l d_{il}}{Z_s^2} e^{j \boldsymbol{\kappa}_i^T \mathbf{z}'} e^{-j \boldsymbol{\kappa}_l^T \mathbf{s}}.
    \end{equation}
    Here, $d_{il}$ is the entry in the $i$-th row and $l$-th column of the matrix $\mathbf{D} = (\mathbf{I}_J + \mathbf{\Lambda} \mathbf{Q})^{-1}$, where $\mathbf{\Lambda} = \mathrm{diag}\{\rho_1/Z_s,\dots,\rho_J/Z_s \}$ and the entry of the matrix $\mathbf{Q}$ in the $i$-th row and $l$-th column is given by 
    \begin{align} \label{Q_function}
        Q_{il} & = \int_{\mathcal{S}} e^{-j (\boldsymbol{\kappa}_i - \boldsymbol{\kappa}_l)^T \mathbf{s}} d \mathbf{s} \nonumber \\
        & = \int_{-\frac{L_x}{2}}^{\frac{L_x}{2}} e^{-j \Delta \kappa_{x, il} s_x} d s_x \int_{-\frac{L_y}{2}}^{\frac{L_y}{2}} e^{-j \Delta \kappa_{y, il} s_y} d s_y \nonumber \\
        & = L_x L_y \mathrm{sinc}\left( \frac{\Delta \kappa_{x,il} L_x}{2} \right) \mathrm{sinc}\left( \frac{\Delta  \kappa_{y,il} L_y}{2} \right),
    \end{align}     
    where $\Delta \kappa_{x,il}$ and $\Delta \kappa_{y,il}$ are the $x$ and $y$ components of $(\boldsymbol{\kappa}_i - \boldsymbol{\kappa}_l)$, respectively, and $\mathrm{sinc}(t) \triangleq \sin(t)/t$.     
\end{proposition}

\begin{IEEEproof}
    The approximated kernel \eqref{approx_overall_kernel} admits the invertible structure described in \cite[Lemma 2]{10938678}. Consequently, the closed-form inverse provided therein can be directly applied to obtain \eqref{inverse_kernel}, which completes the proof.
\end{IEEEproof}

\subsubsection{Closed-form Optimal Beamforming}

Using \textbf{Proposition~\ref{proposition_0}}, we can derive an explicit expression for the optimal beamformer. First, substituting \eqref{inverse_kernel} into \eqref{optimal_inverse_kernel_solution} yields the solution for $v(\mathbf{s})$ as
\begin{align} \label{approximated_v}
    v(\mathbf{s}) & = \frac{1}{Z_s} h^*(\mathbf{s}) - \sum_{i=1}^J \sum_{l=1}^J \frac{\rho_l d_{il}}{Z_s^2} e^{j \boldsymbol{\kappa}_i^T \mathbf{s}} \int_{\mathcal{S}} h^*(\mathbf{z}) e^{-j \boldsymbol{\kappa}_l^T \mathbf{z}} d \mathbf{z} \nonumber \\
    & = \frac{1}{Z_s} h^*(\mathbf{s}) - \frac{1}{Z_s} \sum_{i=1}^J b_i e^{j \boldsymbol{\kappa}_i^T \mathbf{s}}.
\end{align} 
Here, $b_i$ is the $i$-th entry of the vector $\mathbf{b} = \mathbf{D} \mathbf{\Lambda} \mathbf{a}$, where the $l$-th entry of the vector $\mathbf{a}$ is given by 
\begin{align} \label{double_sinc}
    a_l & = \int_{\mathcal{S}} h^*(\mathbf{z}) e^{-j \boldsymbol{\kappa}_l^T \mathbf{z}} d \mathbf{z} = \int_{\mathcal{S}} \beta_{\mathrm{r}}^*  e^{-j (\boldsymbol{\kappa}_l - \boldsymbol{\kappa}_{\mathrm{r}})^T \mathbf{z}} d \mathbf{z} \nonumber \\
    & = \beta_{\mathrm{r}}^* L_x L_y \mathrm{sinc}\left( \frac{\Delta \widetilde{\kappa}_{x,l} L_x}{2} \right) \mathrm{sinc}\left( \frac{\Delta  \widetilde{\kappa}_{y,l} L_y}{2} \right),
\end{align}  
with $\Delta \widetilde{\kappa}_{x,l}$ and $\Delta \widetilde{\kappa}_{y,l}$ representing the respective $x$- and $y$-components of $(\boldsymbol{\kappa}_l - \boldsymbol{\kappa}_{\mathrm{r}})$.
By substituting \eqref{approximated_v} into \eqref{optimal_solution_SU}, we obtain the closed-form optimal beamformer with mutual coupling as
\begin{align}
    w_{\mathrm{opt}}(\mathbf{s}) = \sqrt{ \frac{2 P_{\mathrm{t}}}{Z_s(\eta - \mathbf{a}^H \mathbf{D} \mathbf{\Lambda} \mathbf{a})} } \left(  h^*(\mathbf{s}) -  \sum_{i=1}^J b_i e^{j \boldsymbol{\kappa}_i^T \mathbf{s}} \right),
\end{align}
where $\eta = \int_{\mathcal{S}} |h(\mathbf{s})|^2 d \mathbf{s} = L_x L_y |\beta_{\mathrm{r}}|^2$. Finally, the resulting optimal array gain is given by 
\begin{align}
    G_{\mathrm{opt}} & = \frac{1}{P_{\mathrm{t}}} \left| \int_{\mathcal{S}} h(\mathbf{s}) w_{\mathrm{opt}}(\mathbf{s}) d \mathbf{s} \right|^2 =  \frac{2}{Z_s} (\eta - \mathbf{a}^H \mathbf{D} \mathbf{\Lambda} \mathbf{a}).
\end{align}

It can be observed that the optimal array gain is upper-bounded by $G_{\mathrm{opt}} \le 2 \eta / Z_s$, which essentially represents the ideal case where the radiation mutual coupling term in the power constraint is omitted. All effects of mutual coupling are thus encapsulated in the penalty term $\mathbf{a}^H \mathbf{D} \mathbf{\Lambda} \mathbf{a}$.

\subsection{Conjugate Gradient}

In contrast to the kernel approximation method, the conjugate gradient method aims to solve the Fredholm integral equation in \eqref{Fredholm_equation} without directly inverting the operator. This method reframes the integral equation as a functional optimization problem, as established in the following proposition.

\begin{proposition} \label{proposition_1}
    \normalfont 

    The solution to the Fredholm integral equation in \eqref{Fredholm_equation} is identical to the solution of the quadratic functional optimization problem
    \begin{align} \label{CG_problem}
        \min_{v(\mathbf{s})} \quad \mathcal{J}(v) = \frac{1}{2} \int_{\mathcal{S}} \int_{\mathcal{S}} v^* (\mathbf{s})c(\mathbf{s}-\mathbf{z}) v(\mathbf{z}) d \mathbf{z} d \mathbf{s} \nonumber \\  - \Re \left\{ \int_{\mathcal{S}} h(\mathbf{s}) v(\mathbf{s}) d \mathbf{s} \right\}.
    \end{align} 
\end{proposition}

\begin{IEEEproof}
    Following the principles of the calculus of variations, the optimal solution to \eqref{CG_problem} is found where the first variation of the functional  $\mathcal{J}(v)$, denoted $\delta \mathcal{J}(v, \delta v)$, is zero for any perturbation $\delta v$. The first variation is given by
    \begin{equation} \label{first_variation}
        \delta \mathcal{J}(v, \delta v)  = \left. \frac{d}{d\epsilon} \mathcal{J}(v + \epsilon \delta v) \right|_{\epsilon = 0}
         = - \Re \left\{\int_{\mathcal{S}} \delta v^*(\mathbf{s}) r (\mathbf{s}) d \mathbf{s}  \right\},
    \end{equation}
    where $r(\mathbf{s})$ is the residual defined as
    \begin{equation} \label{residual}
        r(\mathbf{s}) = h^*(\mathbf{s}) - \int_{\mathcal{S}} c(\mathbf{s} - \mathbf{z}) v(\mathbf{z}) d \mathbf{z}.
    \end{equation} 
    For the functional to be at a minimum, we must have $\delta \mathcal{J}(v, \delta v) = 0$ for any arbitrary $\delta v(\mathbf{s})$. This condition implies that the residual $r(\mathbf{s})$ must be zero, which is equivalent to the original Fredholm equation \eqref{Fredholm_equation}. This completes the proof. 
\end{IEEEproof}

\subsubsection{Algorithm Flow}
Based on the equivalence established in \textbf{Proposition~\ref{proposition_1}}, we can solve \eqref{CG_problem} using the conjugate gradient method. From the first variation in \eqref{first_variation}, we identify the gradient of the functional $\mathcal{J}(v)$ as
\begin{equation}
    \nabla \mathcal{J}(v)(\mathbf{s}) = - r(\mathbf{s}).
\end{equation} 
Then, following the standard conjugate gradient procedure for minimization \cite{shewchuk1994introduction}, adapted for the considered continuous functional problem \eqref{CG_problem}, the iterative algorithm is defined by
\begin{align}
    & v_{(n+1)}(\mathbf{s}) = v_{(n)}(\mathbf{s}) + \alpha_{(n)} p_{(n)} (\mathbf{s}), \\
    & p_{(n+1)}(\mathbf{s}) = r_{(n+1)}(\mathbf{s}) + \xi_{(n+1)} p_{(n)} (\mathbf{s}),
\end{align}
where $v_{(n)}$ is the solution estimate, $p_{(n)}$ is the conjugate search direction, and $r_{(n)}$ is the residual at the $n$-th iteration.

The step sizes $\alpha_{(n)}$ and $\xi_{(n)}$ are chosen to optimize the minimization at each step. Following \eqref{residual}, the residual can be updated efficiently using the result from the previous iteration:
\begin{align} \label{update_r}
    & r_{(n+1)}(\mathbf{s}) = r_{(n)}(\mathbf{s}) - \alpha_{(n)} \int_{\mathcal{S}} c(\mathbf{s} - \mathbf{z}) p_{(n)}(\mathbf{z}) d \mathbf{z} \nonumber \\
    & = r_{(n)}(\mathbf{s}) - \alpha_{(n)} \left( \int_{\mathcal{S}} c_{\mathrm{rad}}(\mathbf{s}-\mathbf{z}) p_{(n)}(\mathbf{z}) d \mathbf{z} +  Z_s p_{(n)}(\mathbf{s})  \right).
\end{align} 
The optimal step sizes are derived in the same manner as in~\cite{shewchuk1994introduction} and are given by
\begin{align}
    \alpha_{(n)} & = \frac{\int_{\mathcal{S}} \left| r_{(n)}(\mathbf{s}) \right|^2 d \mathbf{s}}{\int_{\mathcal{S}} \int_{\mathcal{S}} p_{(n)}^* (\mathbf{s})c(\mathbf{s}-\mathbf{z}) p_{(n)}(\mathbf{z}) d \mathbf{z} d \mathbf{s}}, \\
    \xi_{(n+1)} & = \frac{\int_{\mathcal{S}} \left| r_{(n+1)}(\mathbf{s}) \right|^2 d \mathbf{s} }{\int_{\mathcal{S}} \left| r_{(n)}(\mathbf{s}) \right|^2 d \mathbf{s}}.
\end{align}

\subsubsection{Numerical Implementation}
While the algorithm is fully described in its continuous form, its numerical implementation requires discretization. We achieve this using Gauss-Legendre quadrature to approximate the surface integrals over the rectangular aperture $\mathcal{S}$ as
\begin{align}
    \int_{\mathcal{S}} g(\mathbf{s}) d \mathbf{s} & = \int_{-\frac{L_x}{2}}^{\frac{L_x}{2}} \int_{-\frac{L_y}{2}}^{\frac{L_y}{2}} g (\mathbf{s}) d s_x d s_y \nonumber \\
    & \approx \sum_{n=1}^M \sum_{m=1}^M \frac{\omega_n \omega_m L_x L_y}{4} g(\mathbf{s}_{n,m}),
\end{align} 
where the sampled point is $\mathbf{s}_{n,m} = \left[ L_x\theta_n/2, L_y \theta_m/2, 0 \right]^T$. This discretization transforms the continuous functions into vectors evaluated at the grid points $\mathbf{s}_{n,m}$. At these points, the primary conjugate gradient iterations become
\begin{align} \label{CG_iteration_GL}
    & \mathbf{v}_{(n+1)} = \mathbf{v}_{(n)} + \alpha_{(n)} \mathbf{p}_{(n)},  \\
    & \mathbf{p}_{(n+1)} = \mathbf{r}_{(n+1)} + \xi_{(n+1)} \mathbf{p}_{(n)},
\end{align}
where the vectors $\mathbf{v}_{(n)}$, $\mathbf{p}_{(n)}$, and $\mathbf{r}_{(n)}$ contain the function values at the $M \times M$ sample points, e.g., $\mathbf{v}_{(n)} = [v_{(n)}(\mathbf{s}_{1,1}), \dots, v_{(n)}(\mathbf{s}_{M,M}) ]^T$.

Let $\mathbf{z}_{n,m} = \mathbf{s}_{n,m}$ for notational convenience. Then, by using the Gauss-Legendre quadrature, the residual update in \eqref{update_r} can be implemented as
\begin{align}
    \mathbf{r}_{(n+1)} & = \mathbf{r}_{(n)} - \alpha_{(n)} \left( \mathbf{C}_{\mathrm{rad}} \mathbf{\Phi}  \mathbf{p}_{(n)} + Z_s \mathbf{p}_{(n)} \right),
\end{align} 
where $\mathbf{\Phi}$ is a diagonal matrix of the quadrature weights and $\mathbf{C}_{\mathrm{rad}}$ is the matrix of the radiation mutual coupling kernel evaluated at all pairs of sample points  
\begin{align}
    & \mathbf{\Phi} = \frac{L_x L_y}{4} \mathrm{diag} \left(\omega_1 \omega_1,\dots,\omega_M \omega_M  \right), \\
    & \mathbf{C}_{\mathrm{rad}} = \begin{bmatrix}
        c_{\mathrm{rad}} (\mathbf{s}_{1,1} - \mathbf{z}_{1,1}) & \cdots & c_{\mathrm{rad}} (\mathbf{s}_{1,1} - \mathbf{z}_{M,M}) \\
        \vdots & \ddots & \vdots \\
        c_{\mathrm{rad}} (\mathbf{s}_{M,M} - \mathbf{z}_{1,1}) & \cdots & c_{\mathrm{rad}} (\mathbf{s}_{M,M} - \mathbf{z}_{M,M})
    \end{bmatrix}.
\end{align}
It is worth noting that for an arbitrary initialization $\mathbf{v}_{(0)}$, the corresponding residual vector $\mathbf{r}_{(0)}$ is calculated directly from its definition in \eqref{residual} as
\begin{equation}
    \mathbf{r}_{(0)} = \mathbf{h}^* - \left( \mathbf{C}_{\mathrm{rad}} \mathbf{\Phi}  \mathbf{v}_{(0)} + Z_s \mathbf{v}_{(0)} \right),
\end{equation} 
where $\mathbf{h} = [h(\mathbf{s}_{1,1}), \dots, h(\mathbf{s}_{M,M}) ]^T$. Similarly, the step sizes can be numerically calculated by 
\begin{align}
    \alpha_{(n)} & = \frac{\mathbf{r}_{(n)}^H \mathbf{\Phi} \mathbf{r}_{(n)}}{\mathbf{p}_{(n)}^H \mathbf{\Phi} \mathbf{C}_{\mathrm{rad}} \mathbf{\Phi} \mathbf{p}_{(n)} + Z_s \mathbf{p}_{(n)}^H \mathbf{\Phi} \mathbf{p}_{(n)} },  \\
    \xi_{(n+1)} & = \frac{\mathbf{r}_{(n+1)}^H \mathbf{\Phi} \mathbf{r}_{(n+1)}}{\mathbf{r}_{(n)}^H \mathbf{\Phi} \mathbf{r}_{(n)}}.
\end{align}

Once the conjugate gradient iteration \eqref{CG_iteration_GL} converges to a solution vector $\mathbf{v}_{\mathrm{opt}}$, the continuous solution $v(\mathbf{s})$ can be reconstructed for any point $\mathbf{s} \in \mathcal{S}$ based on \eqref{Fredholm_equation} as 
\begin{equation}
    v_{\mathrm{opt}}(\mathbf{s}) = \frac{1}{Z_s} \left( h^*(\mathbf{s}) - \mathbf{c}_{\mathrm{rad}}^T(\mathbf{s}) \mathbf{\Phi} \mathbf{v}_{\mathrm{opt}} \right),
\end{equation} 
where $\mathbf{c}_{\mathrm{rad}}(\mathbf{s}) = [c_{\mathrm{rad}}(\mathbf{s} - \mathbf{z}_{1,1}), \dots, c_{\mathrm{rad}}(\mathbf{s} - \mathbf{z}_{M,M})]^T$. Finally, the optimal beamformer is synthesized from this continuous solution following \eqref{optimal_solution_SU} as
\begin{equation}
    w_{\mathrm{opt}}(\mathbf{s}) = \sqrt{\frac{2 P_{\mathrm{t}}}{\mathbf{h}^T \mathbf{\Phi} \mathbf{v}_{\mathrm{opt}}}} v_{\mathrm{opt}}(\mathbf{s}).
\end{equation}

\begin{remark}
    \normalfont

    \emph{(Applicability to Near-Field Channels)} While this work primarily evaluates beamforming under the far-field channel approximation in \eqref{far_field_channel}, the proposed optimization frameworks can be discussed in the context of near-field communications. Specifically, the conjugate gradient method is inherently channel-agnostic and can be directly applied to the exact near-field spherical wave channel model formulated in \eqref{near_field_channel}. This requires no algorithmic modifications. Conversely, the KA method leverages the far-field plane-wave assumption to analytically evaluate the spatial integrals as a product of $\mathrm{sinc}$ functions, yielding the closed-form solutions in \eqref{Q_function} and \eqref{double_sinc}. If applied to the near-field, this specific spatial integral would lose its closed-form tractability, necessitating numerical integration, such as Gauss-Legendre quadrature, to compute the optimal continuous beamformer.
\end{remark}

\subsubsection{Convergence}

We briefly discuss the convergence behavior of the proposed conjugate gradient method. After Gauss-Legendre discretization, solving the continuous optimization problem in \eqref{CG_problem} can be viewed as solving the following linear system:
\begin{equation}
    \mathbf{A}\mathbf{v}=\mathbf{h}^*, \quad
    \mathbf{A}=\mathbf{C}_{\mathrm{rad}}\mathbf{\Phi}+Z_s\mathbf{I},
\end{equation}
where $\mathbf{\Phi}$ is a diagonal matrix containing the positive quadrature weights. To analyze this system, we introduce the change of variables $\bar{\mathbf{v}}=\mathbf{\Phi}^{1/2}\mathbf{v}$ and $\bar{\mathbf{h}}=\mathbf{\Phi}^{1/2}\mathbf{h}^*$, which leads to the following equivalent transformed system:
\begin{equation}
    \bar{\mathbf{A}}\bar{\mathbf{v}}=\bar{\mathbf{h}}, \quad
    \bar{\mathbf{A}}=\mathbf{\Phi}^{1/2}\mathbf{A}\mathbf{\Phi}^{-1/2}
    =\mathbf{\Phi}^{1/2}\mathbf{C}_{\mathrm{rad}}\mathbf{\Phi}^{1/2}+Z_s\mathbf{I}.
\end{equation}
Since $\mathbf{\Phi}^{1/2}$ is invertible, the transformed system is fully equivalent to the original one. Moreover, from the wavenumber-domain representation in \eqref{WD_coupling_kernel}, the radiation kernel has a nonnegative spectrum, which implies that $\mathbf{C}_{\mathrm{rad}}$ is Hermitian positive semidefinite. Hence $\mathbf{\Phi}^{1/2}\mathbf{C}_{\mathrm{rad}}\mathbf{\Phi}^{1/2}$ is also Hermitian positive semidefinite. Since the dissipation term satisfies $Z_s>0$, the matrix $\bar{\mathbf{A}}$ is Hermitian positive definite, and the discretized objective is therefore strictly convex with a unique minimizer. It then follows from the classical conjugate gradient theory that the proposed iteration converges to a unique solution for any initialization \cite{shewchuk1994introduction}. 
In addition, the convergence speed is mainly governed by the condition number of $\bar{\mathbf{A}}$. In practice, a larger quadrature order $M$ increases the problem dimension, and a higher operating frequency makes the coupling kernel more oscillatory. Both effects may worsen the conditioning of the discretized system and therefore slow down the conjugate gradient iterations.

\section{Array Directivity and Beampattern Analysis}

\subsection{Array Directivity Analysis} \label{sec_array_gain}

This subsection further explores the impact of mutual coupling on optimal array gain, with an emphasis on the impact on directivity. To this end, we begin by providing the following proposition for the array gain.

\begin{proposition} \label{proposition_infinite_array_gain}
    \normalfont

    For an infinite transmit aperture, i.e., $|\mathcal{S}| \rightarrow +\infty$, the optimal normalized array gain per unit area at the receiver with mutual coupling under the far-field assumption is given by 
    \begin{align} \label{optimal_array_gain_infinite}
        \lim_{|\mathcal{S}| \rightarrow +\infty} \frac{G_{\mathrm{opt}}}{|\mathcal{S}|} & = \frac{2|\beta_{\mathrm{r}}|^2}{Z_s + C_{\mathrm{rad}}(\boldsymbol{\kappa}_{\mathrm{r}})} = \frac{1}{4\pi^2} \left(\frac{\kappa_0}{R_0}\right)^2 D(\theta_0, \phi_0),
    \end{align} 
    where
    \begin{equation}
        D(\theta, \phi) = \frac{Z_0^2(1 - \sin^2 \theta \sin^2 \phi)^2 \cos \phi}{2 Z_s \cos \phi + Z_0(1 - \sin^2\theta \sin^2 \phi)}.
    \end{equation} 
\end{proposition}

The result in \textbf{Proposition \ref{proposition_infinite_array_gain}} shows that, in the large-aperture limit, the optimal array gain scales linearly with the aperture area, and its angular dependency is captured by the directivity factor $D(\theta, \phi)$. This allows us to focus on the impact of polarization and mutual coupling without being concerned with the absolute aperture size. Specifically, if there is no mutual coupling, i.e., $C_{\mathrm{rad}}(\boldsymbol{\kappa}_{\mathrm{r}}) = 0$, the directivity of the array gain is merely determined by $|\beta_{\mathrm{r}}|^2$, which is captured by the polarization term $\left( 1 - \sin^2 \theta_0 \sin^2 \phi_0 \right)^2$. When mutual coupling exists, the directivity is also determined by the value of the wavenumber-domain mutual coupling kernel at the receiver direction $(\theta_0, \phi_0)$. In the following, we focus on the directivity in the E-Plane (i.e., $y$-$z$ plane aligned with the polarization direction) and the H-Plane (i.e., $x$-$z$ plane orthogonal to the polarization direction), respectively.
\begin{itemize}
    \item \textbf{E-Plane} ($\theta = \pi/2$): In this plane, the directivity factor $D(\theta, \phi)$ reduces to 
    \begin{equation} \label{E_plane}
        D_{\mathrm{E}}(\phi) = D(\pi/2, \phi) = \frac{Z_0^2 \cos^4 \phi}{2Z_s + Z_0 \cos \phi}. 
    \end{equation}
    
    \item \textbf{H-Plane} ($\theta = 0$): In this plane, the directivity factor $D(\theta, \phi)$ simplifies to 
    \begin{equation} \label{H_plane}
        D_{\mathrm{H}}(\phi) = D(0, \phi) = \frac{Z_0^2 \cos \phi}{2 Z_s \cos \phi + Z_0}.
    \end{equation}
\end{itemize}

The above results provide the following insights. 1) The array gain is maximized at front-fire ($\phi = 0$), where the directivity factor reaches its peak value of $D(\theta, 0) = Z_0^2 / (2Z_s + Z_0)$. As the angle approaches the end-fire ($\phi \rightarrow \pm \pi/2$), the $\cos\phi$ term in the numerator forces the gain to zero. 2) The array gain is anisotropic, showing a clear dependence on the angle $\theta$ due to $y$-polarization. For any angle $\phi$ off front-fire, the $\cos^4\phi$ term in the E-plane formula \eqref{E_plane} will be significantly smaller than the $\cos\phi$ term in the H-plane formula \eqref{H_plane}. This shows that the array gain is stronger in the H-plane and drops off much more quickly in the E-plane as the angle $\phi$ increases.

However, a simplified model that omits mutual coupling and assumes the gain is determined solely by the channel response with polarization projection, i.e., $|\beta|^2 \propto (1 - \sin^2\theta \sin^2\phi)^2$, fails to capture this complete physical picture. Specifically, in the E-plane where $\theta = \pi/2$, this simple model yields a gain proportional to $(1 - \sin^2\phi)^2 = \cos^4\phi$. This coincidentally predicts that the gain vanishes at the end-fire. In the H-plane where $\theta = 0$, the array gain becomes proportional to $(1 - 0)^2 = 1$. This incorrectly implies the gain is constant for all $\phi$. 

It is important to note that the above analysis is accurate only in the large-aperture limit. For practical small apertures, the spatial windowing effect from the finite aperture may lead to significant power leakage at the end-fire ($\phi = \pm \pi/2$). This leakage is particularly relevant in the H-plane ($\theta = 0$), where the polarization projection factor $\beta$ does not vanish at the end-fire. Such effects are non-trivial to characterize theoretically and need to be accurately evaluated using the numerical methods proposed in Section \ref{optimization_method}.

\subsection{Beampattern Analysis} \label{sec_beampattern}

We now focus on the beampattern, which is defined as the electric field strength received by a $y$-polarized receiver in any given far-field direction. For any projected wavevector $\boldsymbol{\kappa}
= \kappa_0[\cos\theta\sin\phi,\sin\theta\sin\phi,0]^T$ and any given beamformer $w(\mathbf{s})$, the beampattern is given by
\begin{equation}
    B(\boldsymbol{\kappa}) =  \left| \int_{\mathcal{S}} \hat{\beta}(\boldsymbol{\kappa}) w(\mathbf{s}) e^{-j \boldsymbol{\kappa}^T \mathbf{s}} d \mathbf{s} \right|
    =  \left| \hat{\beta}(\boldsymbol{\kappa}) \hat{W}(\boldsymbol{\kappa}) \right|.
\end{equation} 
Here, $\hat{\beta}(\boldsymbol{\kappa}) = \left( 1 - \sin^2 \theta \sin^2 \phi \right)$ is the polarization projection factor as in \eqref{beta_factor} and $\hat{W}(\boldsymbol{\kappa})$ is the windowed 2D Fourier transform of $w(\mathbf{s})$, defined as
\begin{equation}
    \hat{W}(\boldsymbol{\kappa}) = \mathcal{F}_{\mathcal{S}}\{w\}(\boldsymbol{\kappa}) = \int_{\mathcal{S}} w(\mathbf{s}) e^{-j \boldsymbol{\kappa}^T \mathbf{s}} d \mathbf{s}.
\end{equation} 
We next discuss the beampatterns for the uncoupled and coupled cases, respectively.

\subsubsection{Uncoupled CAPA}
For uncoupled CAPAs, the optimal beamformer is the matched filtering solution, i.e., $w_{\mathrm{opt}}(\mathbf{s}) = \gamma h^*(\mathbf{s})$, where $\gamma$ is a scaling factor to satisfy the uncoupled power constraint discussed in \textbf{Remark \ref{remark_uncoupled}}. Following \eqref{double_sinc}, $\hat{W}(\boldsymbol{\kappa})$ for uncoupled CAPAs can be calculated as
\begin{align}
    \hat{W}_{\text{uncoupled}}(\boldsymbol{\kappa}) & = \gamma \int_{\mathcal{S}} h^*(\mathbf{s}) e^{-j \boldsymbol{\kappa}^T \mathbf{s}} d \mathbf{s} \nonumber \\
    & \hspace{-1cm} = \gamma \beta_{\mathrm{r}}^* L_x L_y \mathrm{sinc}\left( \frac{\Delta \bar{\kappa}_{x} L_x}{2} \right) \mathrm{sinc}\left( \frac{\Delta  \bar{\kappa}_{y} L_y}{2} \right) \nonumber \\
    & \hspace{-1cm} = \gamma \hat{H}(\boldsymbol{\kappa}),
\end{align} 
where $\Delta \bar{\kappa}_{x}$ and $\Delta \bar{\kappa}_{y}$ are the $x$ and $y$ components of $(\boldsymbol{\kappa} - \boldsymbol{\kappa}_{\mathrm{r}})$, and $\hat{H}(\boldsymbol{\kappa})$ is essentially the windowed 2D Fourier transform of $h^*(\mathbf{s})$. Consequently, the uncoupled beampattern is
\begin{equation}
    B_{\text{uncoupled}}(\boldsymbol{\kappa}) = \big|\gamma \hat{\beta}(\boldsymbol{\kappa}) \hat{H}(\boldsymbol{\kappa}) \big|.
\end{equation}

\subsubsection{Coupled CAPA} 

Based on \textbf{Theorem \ref{theorem_1}}, the optimal beamformer for a coupled CAPA is proportional to $v(\mathbf{s})$, i.e., $w_{\mathrm{opt}}(\mathbf{s}) = \widetilde{\gamma} v(\mathbf{s})$, where $\widetilde{\gamma}$ is a scaling factor to fulfill the coupled power constraint. The integral equation \eqref{Fredholm_equation} is a spatial convolution over the finite aperture $\mathcal{S}$. By assuming the aperture is large, the convolution theorem can be applied to approximate this relationship in the wavenumber domain as
\begin{align}
    C(\boldsymbol{\kappa}) \cdot \mathcal{F}_{\mathcal{S}} \{v\}(\boldsymbol{\kappa}) \approx \hat{H}(\boldsymbol{\kappa}),
\end{align}
where $C(\boldsymbol{\kappa}) = Z_s + C_{\mathrm{rad}}(\boldsymbol{\kappa})$ denotes the Fourier transform of the overall coupling kernel $c(\mathbf{s})$. It is important to note that this approximation is accurate only when the transmit aperture is sufficiently large. For practical finite apertures, this model ignores significant windowing effects. Based on this large-aperture approximation, $\hat{W}(\boldsymbol{\kappa})$ for coupled CAPAs is approximated by
\begin{equation}
    \hat{W}_{\text{coupled}}(\boldsymbol{\kappa}) = \widetilde{\gamma} \mathcal{F}_{\mathcal{S}} \{v\}(\boldsymbol{\kappa}) \approx \frac{\widetilde{\gamma} \hat{H}(\boldsymbol{\kappa})}{Z_s + C_{\mathrm{rad}}(\boldsymbol{\kappa})}.
\end{equation}
Consequently, the approximate coupled beampattern is
\begin{equation}
    B_{\text{coupled}}(\boldsymbol{\kappa}) \approx \left|\frac{\widetilde{\gamma} \hat{\beta}(\boldsymbol{\kappa})  \hat{H}(\boldsymbol{\kappa})}{Z_s + C_{\mathrm{rad}}(\boldsymbol{\kappa})} \right|.
\end{equation}

\subsubsection{Impact of Mutual Coupling}

To isolate the effect of mutual coupling in this approximate model, we define the following ratio between the approximate coupled beampattern and the uncoupled beampattern, given by
\begin{equation}
    S(\boldsymbol{\kappa}) = \frac{B_{\text{coupled}}(\boldsymbol{\kappa})}{B_{\text{uncoupled}}(\boldsymbol{\kappa}) } \approx \frac{1}{\varrho \left( Z_s + C_{\mathrm{rad}}(\boldsymbol{\kappa}) \right)},
\end{equation}
where $\varrho = \gamma/\widetilde{\gamma}$. This ratio illustrates how the beampattern is filtered by the wavenumber-domain coupling kernel $C_{\mathrm{rad}}(\boldsymbol{\kappa})$ in the large-aperture limit, as discussed in \cite{11006094}. Again, for practical small apertures, the significant windowing effect must also be taken into account, which needs to be accurately evaluated using the numerical methods proposed in Section~\ref{optimization_method}.

\section{Extension to Spatially Discrete Arrays} \label{sec:discrete}

An SPDA model can be obtained from the proposed model of CAPAs. We model the SPDA by partitioning the continuous surface $\mathcal{S}$ into $N$ uniform, non-overlapping antenna elements. Let $\mathcal{S}_{n} \subset \mathcal{S}$ denote the surface of the $n$-th antenna, centered at position $\mathbf{p}_n$. We assume that these element surfaces are disjoint, i.e., $\mathcal{S}_n \cap \mathcal{S}_{m} = \emptyset$ for $n \neq m$, and each has an identical area $|\mathcal{S}_n| = A_{\mathrm{d}}$. Furthermore, we first define a reference element subsurface $\mathcal{S}_{\mathrm{d}}$ centered at the origin as
\begin{equation}
    \mathcal{S}_{\mathrm{d}} = \left\{ [s_x, s_y, 0]^T\, \left|\, |s_x| \le \frac{L_{\mathrm{d}, x}}{2}, |s_y| \le \frac{L_{\mathrm{d}, y}}{2} \right. \right\},
\end{equation} 
where $L_{\mathrm{d}, x} L_{\mathrm{d}, y} = A_{\mathrm{d}}$. The surface of the $n$-th antenna is then a simple translation of this reference, i.e., $\mathcal{S}_n = \mathcal{S}_{\mathrm{d}} + \mathbf{p}_n$.

We assume each antenna element exhibits an identical $y$-polarized current profile $a_{\mathrm{t}}(\mathbf{s}) \mathbf{u}_y$. The total $y$-polarized component of the source current $w_{\mathrm{d}}(\mathbf{s})$ for the SPDA is the weighted superposition of all $N$ elements, given by
\begin{equation} \label{discrete_beamformer}
    w_{\mathrm{d}}(\mathbf{s}) = \sum_{n=1}^{N} v_n   \Pi_n(\mathbf{s}) a_{\mathrm{t}} (\mathbf{s} - \mathbf{p}_n),
\end{equation}   
where $v_n \in \mathbb{C}$ is the discrete beamforming coefficient for the $n$-th antenna, $a_{\mathrm{t}} (\mathbf{s} - \mathbf{p}_n)$ is the element's current profile, shifted to its center $\mathbf{p}_n$ and $\Pi_n(\mathbf{s})$ is a rectangular window function that confines the current to the $n$-th element's surface, defined as $\Pi_n(\mathbf{s}) = 1 $ for $\mathbf{s} \in \mathcal{S}_n$ and $\Pi_n(\mathbf{s}) = 0$ otherwise. The CAPA model can be viewed as a continuum idealization of an electrically dense radiating structure. In a practical array, the continuous surface can be realized through finite antenna elements, each occupying a small subsurface and supporting a finite current distribution.

\subsection{Mutual Coupling}

Based on the above model, the transmit EM power for the SPDA is obtained by substituting \eqref{discrete_beamformer} into \eqref{em_power}, yielding
\begin{align} \label{EM_power_discrete}
    P_{\mathrm{em}}^{\mathrm{d}} & =  \frac{1}{2} \int_{\mathcal{S}} \int_{\mathcal{S}} w_{\mathrm{d}}^*(\mathbf{s}) c(\mathbf{s} - \mathbf{z}) w_{\mathrm{d}}(\mathbf{z}) d \mathbf{z} d \mathbf{s} \nonumber \\
    & = \frac{1}{2} \sum_{n=1}^{N} \sum_{m=1}^{N} v_n^* v_m \nonumber \\
    & \hspace{1cm} \times \underbrace{\int_{\mathcal{S}_n} \int_{\mathcal{S}_m} \hspace{-0.2cm} a_{\mathrm{t}}^* (\mathbf{s} - \mathbf{p}_n) c(\mathbf{s}-\mathbf{z}) a_{\mathrm{t}} (\mathbf{z} - \mathbf{p}_m) d \mathbf{z} d \mathbf{s}}_{\triangleq \Psi_{nm}}.
\end{align}
By denoting $\boldsymbol{v} = [v_{1},\dots,v_{N}]^T$, the transmit EM power can be written in the compact matrix form
\begin{equation}
    P_{\mathrm{em}}^{\mathrm{d}} = \frac{1}{2} \boldsymbol{v}^H \mathbf{\Psi} \boldsymbol{v},
\end{equation} 
where the entry of the coupling matrix $\mathbf{\Psi}$ in the $n$-th row and $m$-th column is $\Psi_{nm}$, as defined in \eqref{EM_power_discrete}. Based on \eqref{scalar_coupling}, the matrix $\mathbf{\Psi}$ can be written explicitly as
\begin{equation} \label{SPDA_mutual_coupling_matrix}
    \mathbf{\Psi} = Z_{\mathrm{d}, s} \mathbf{I}_N + \mathbf{\Psi}_{\mathrm{rad}},
\end{equation} 
where $Z_{\mathrm{d},s}$ is the self-impedance of individual antenna elements and $\mathbf{\Psi}_{\mathrm{rad}}$ is the radiation mutual coupling matrix. Specifically, $Z_{\mathrm{d},s}$ is given by 
\begin{align}
    Z_{\mathrm{d},s} = Z_s \int_{\mathcal{S}_{\mathrm{d}}} \left|a_{\mathrm{t}}(\mathbf{s}) \right|^2 d \mathbf{s}.
\end{align}
Furthermore, the $(n,m)$-th entry of $\mathbf{\Psi}_{\mathrm{rad}}$ is given by 
\begin{align} \label{SPDA_coupling_matrix}
   \mathbf{\Psi}_{\mathrm{rad}}(n,m) & \! = \! \int_{\mathcal{S}_n} \! \int_{\mathcal{S}_m} \hspace{-0.2cm} a_{\mathrm{t}}^* (\mathbf{s} - \mathbf{p}_n) c_{\mathrm{rad}}(\mathbf{s}-\mathbf{z}) a_{\mathrm{t}} (\mathbf{z} - \mathbf{p}_m) d \mathbf{z} d \mathbf{s} \nonumber \\
   &\! = \!\int_{\mathcal{S}_{\mathrm{d}}} \! \int_{\mathcal{S}_{\mathrm{d}}} a_{\mathrm{t}}^* (\mathbf{s}) c_{\mathrm{rad}}(\mathbf{s}-\mathbf{z} + \Delta \mathbf{p}_{nm}) a_{\mathrm{t}} (\mathbf{z}) d \mathbf{z} d \mathbf{s},
\end{align}   
where $\Delta \mathbf{p}_{nm} = \mathbf{p}_n - \mathbf{p}_m$ denotes the spacing between the $n$-th and the $m$-th antenna elements.
This construction makes the connection to practical antenna implementations explicit. Particularly, the discrete coupling matrix is obtained by integrating the continuous kernel over the current profiles of the transmitting elements, which is consistent with the mutual-impedance viewpoint used in conventional array modeling, e.g., \cite[Eq. (5)]{10547020}. Moreover, note that the choice of current profile $a_{\mathrm{t}}(\mathbf{s})$ significantly affects the mutual coupling, and it is possible to design $a_{\mathrm{t}}(\mathbf{s})$ carefully to minimize the mutual coupling between different elements. 

Assuming a slowly changing current profile $a_{\mathrm{t}}(\mathbf{s})$ over the reference surface $\mathcal{S}_{\mathrm{d}}$ and a small aperture for each antenna element, the radiation mutual coupling matrix can be approximated as
\begin{align}
     \mathbf{\Psi}_{\mathrm{rad}}(n,m) & \approx A_{\mathrm{d}}^2 \left|a_{\mathrm{t}} (\mathbf{0})\right|^2 c_{\mathrm{rad}}(\Delta \mathbf{p}_{nm}).
\end{align}

Based on the analysis in \textbf{Proposition \ref{proposition_null}}, a half-wavelength spacing cannot null the mutual coupling for polarized antennas, i.e., $c_{\mathrm{rad}}(\Delta \mathbf{p}_{nm}) \neq 0$ for $\|\Delta \mathbf{p}_{nm}\| = \lambda/2$. This is in contrast with the case of ideal isotropic antennas that are widely assumed in the literature.

The above SPDA model captures several practical finite-array effects, including finite aperture, finite element area, finite inter-element spacing, and element current profiles. At the same time, it remains a reduced analytical model rather than a geometry-specific full-wave representation of a particular antenna implementation. In particular, effects associated with detailed element shape, feeding structure, and matching networks are not explicitly modeled here. Instead, the model provides a general framework for understanding the fundamental mutual coupling effects in practical arrays and can be used to guide the design of array geometries and element current profiles to mitigate coupling.

\begin{figure*}[t!]
    \centering
    \begin{subfigure}[t]{0.245\textwidth}
        \includegraphics[width=1\textwidth]{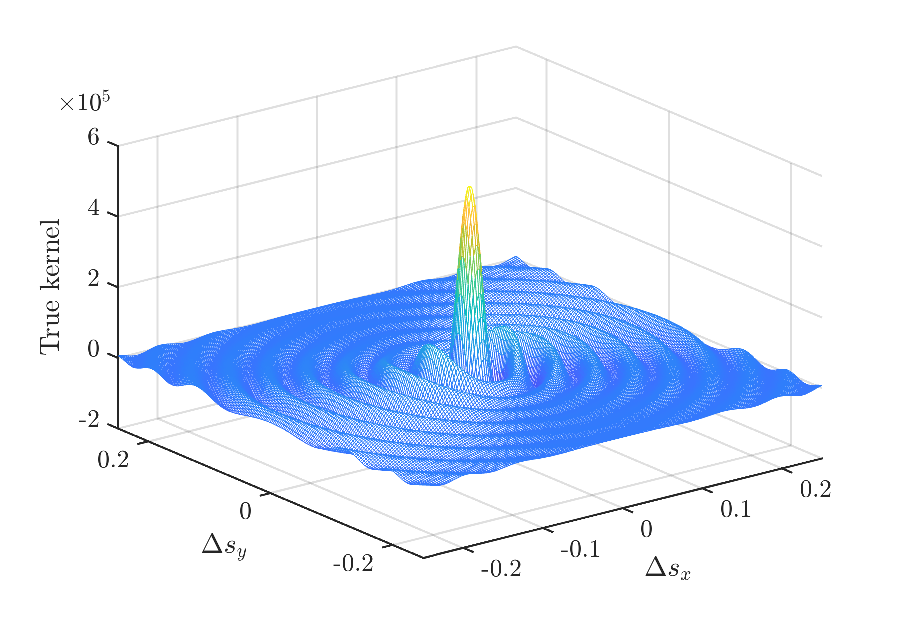}
        \caption{True kernel.}
    \end{subfigure}
    \begin{subfigure}[t]{0.245\textwidth}
        \includegraphics[width=1\textwidth]{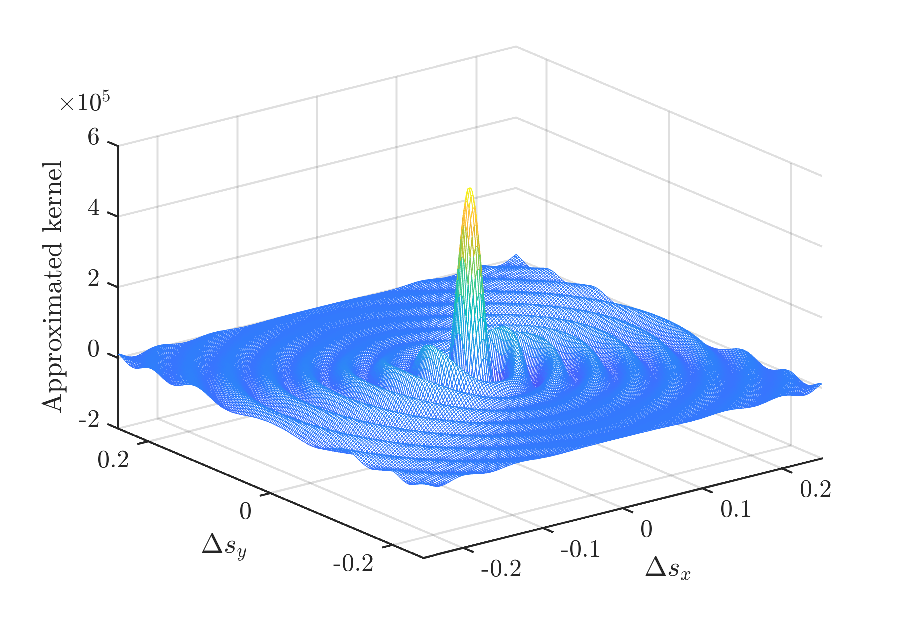}
        \caption{Approximated, $M = 30$.}
    \end{subfigure}
        \begin{subfigure}[t]{0.245\textwidth}
        \includegraphics[width=1\textwidth]{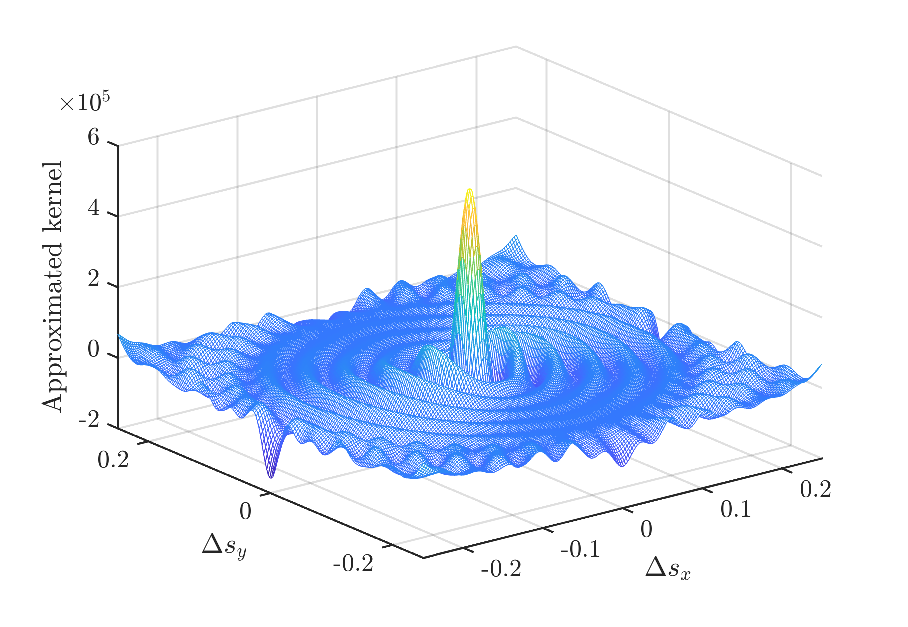}
        \caption{Approximated, $M = 20$.}
    \end{subfigure}
    \begin{subfigure}[t]{0.245\textwidth}
        \includegraphics[width=1\textwidth]{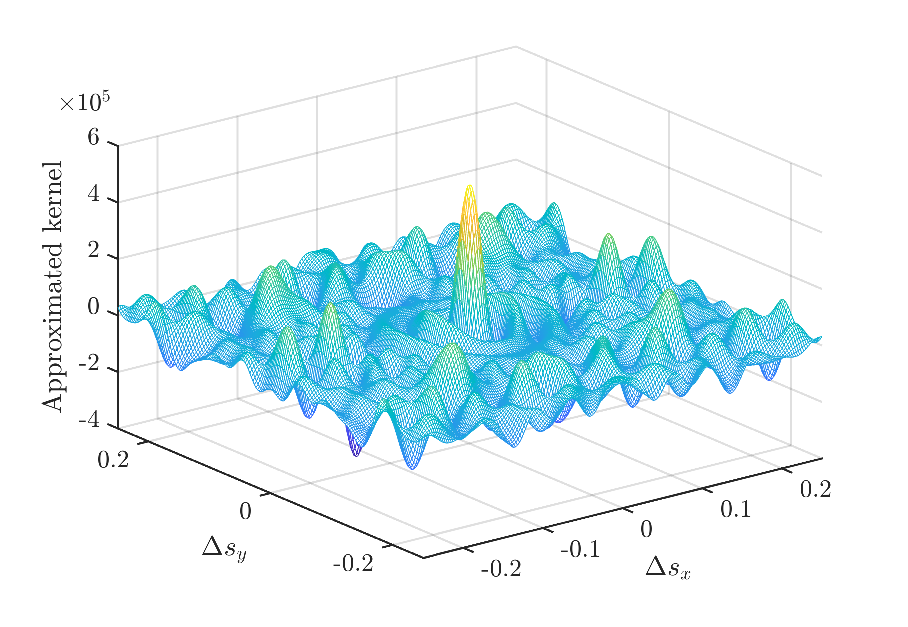}
        \caption{Approximated, $M = 10$.}
    \end{subfigure}
    \caption{Comparison of the true radiation mutual coupling kernel \eqref{radiation_mutual_coupling_kernel} and its approximation in \eqref{proposed_approx_kernel} at $7.8$ GHz for different Gauss-Legendre quadrature orders $M$. As $M$ increases, the approximated kernel more accurately reproduces the shape of the true kernel.}
    \label{fig_approx_accuracy}
    \vspace{-0.4cm}
\end{figure*} 

\subsection{Beamforming}

The electric field at the receiver for discrete arrays is obtained by substituting \eqref{discrete_beamformer} into \eqref{receive_electric_field}, leading to
\begin{align} \label{SPDA_signal_model}
    e_{\mathrm{d}, \mathrm{r}} & = \int_{\mathcal{S}} h(\mathbf{s}) w_{\mathrm{d}}(\mathbf{s}) d \mathbf{s} \nonumber \\
    & = \sum_{n=1}^N v_n \underbrace{\int_{\mathcal{S}_n} h(\mathbf{s}) a_{\mathrm{t}}(\mathbf{s} - \mathbf{p}_n) d \mathbf{s}}_{h_n} = \bm{h}^H \bm{v},
\end{align}
where $\bm{h} = [h_1,\dots,h_N]^H$ is the channel vector for the SPDA, and $\bm{v} = [v_1,\dots,v_N]^T$ is the beamformer vector. Consequently, the beamforming optimization problem for the SPDA can be formulated as 
\begin{subequations}
    \begin{align}
        \max_{\bm{v}} & \quad \left| \bm{h}^H \bm{v} \right|^2 \\
        \mathrm{s.t.} & \quad \frac{1}{2} \bm{v}^H \mathbf{\Psi} \bm{v} \le P_{\mathrm{t}}.
    \end{align}
\end{subequations}
The optimal beamformer can be readily obtained through the whitened matched filter as 
\begin{equation} \label{optimal_beamformer_discrete}
    \bm{v}_{\mathrm{opt}} = \sqrt{\frac{2 P_{\mathrm{t}}}{\bm{h}^H \mathbf{\Psi}^{-1} \bm{h}}} \mathbf{\Psi}^{-1} \bm{h}.
\end{equation}
The resulting optimal array gain is given by 
\begin{equation}
    G_{\mathrm{opt}}^{\mathrm{d}} = \frac{1}{P_{\mathrm{t}}} \left| \bm{h}^H \bm{v}_{\mathrm{opt}} \right|^2 = 2 \bm{h}^H \mathbf{\Psi}^{-1} \bm{h}.
\end{equation}

The optimal discrete beamformer \eqref{optimal_beamformer_discrete} and the continuous one in \textbf{Theorem \ref{theorem_1}} exhibit similar structures. Both solutions require inverting a coupling term. However, the continuous case involves inverting a kernel, which is significantly more challenging than inverting the finite matrix of the discrete case.

\section{Numerical Results} \label{sec:results}

This section presents numerical examples to validate the proposed mutual coupling kernel approximation and the corresponding beamforming designs. Unless specified otherwise, all simulations share the following setup. Specifically, the signal frequency is $f = 2.4$ GHz. The transmit surface is a square with dimensions $L_x = L_y = 0.5$ m. The receiver is placed in the far-field at a distance of $R_0 = 50$ m, which satisfies the far-field condition $R_0 \ge 2D^2/\lambda = 8$ m, with $D = \sqrt{L_x^2 + L_y^2}$ denoting the aperture of the transmit surface. The transmit surface is modeled as copper, with conductivity $\sigma_s = 5.8 \times 10^7$ S/m and permeability $\mu_s = 4 \pi \times 10^{-7}$ H/m. The Gauss-Legendre order is set to $M = 20$. For SPDAs, we assume a simple uniform current profile $a_{\mathrm{t}}(\mathbf{s}) = 1/\sqrt{A_{\mathrm{d}}}$, an element aperture $0.1\lambda \times 0.1 \lambda$ \cite{9906802, wong2014design}, and an antenna spacing of $\lambda/2$. Unless otherwise stated, all numerical results correspond to the uni-polarized configuration in Section II-B, where the transmit current is aligned with the $y$-axis and the receiver is polarization matched.

\begin{figure}[t]
    \centering
    \includegraphics[width=0.45\textwidth]{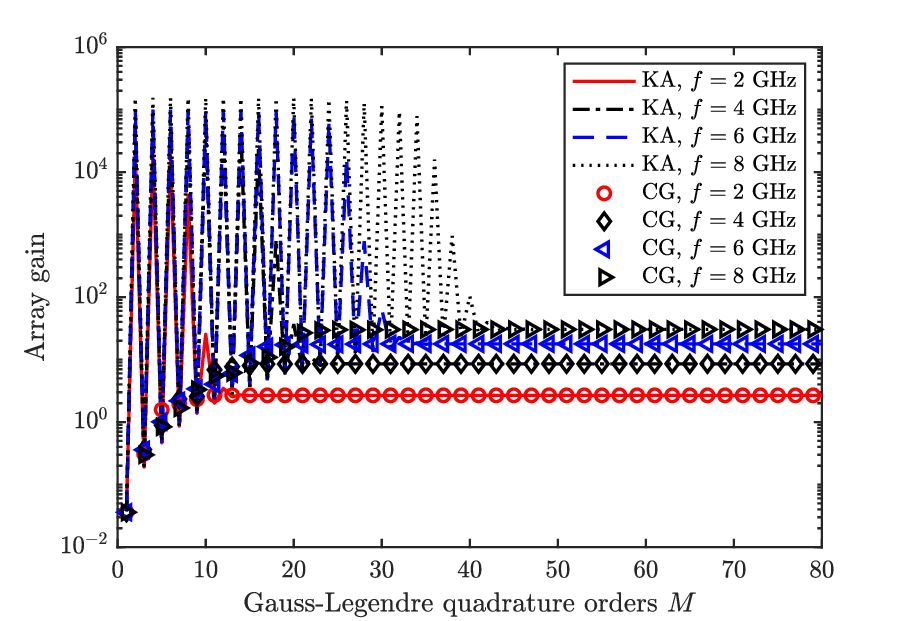}
   \caption{Convergence of the array gain with respect to the Gauss-Legendre quadrature order $M$.}
    \label{fig_GL_convergence}

    \includegraphics[width=0.45\textwidth]{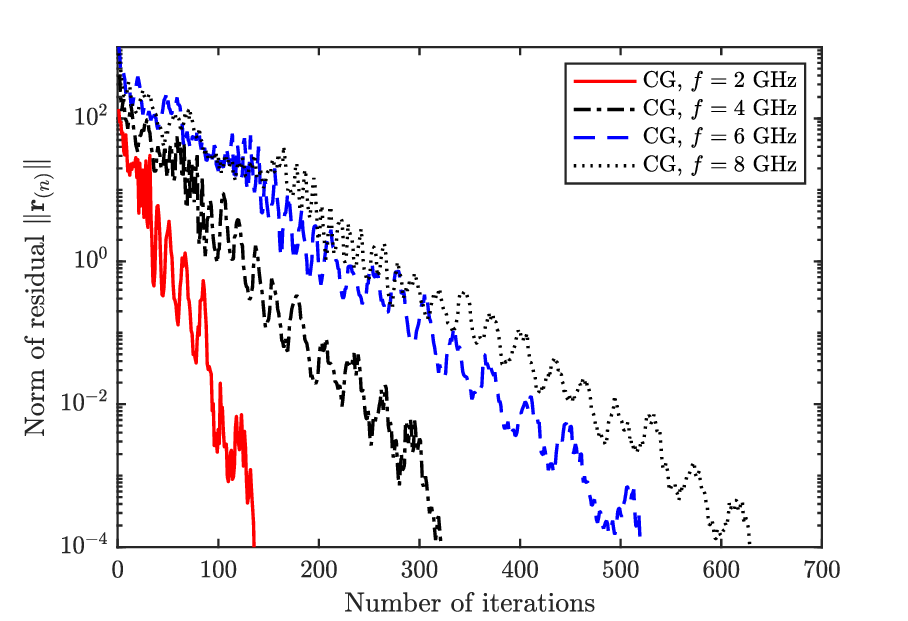}
    \caption{Convergence of the CG method versus the iteration number under different frequencies. }
    \label{fig_CG_convergence}
\end{figure}

\begin{figure}[t]
    \centering
    \includegraphics[width=0.45\textwidth]{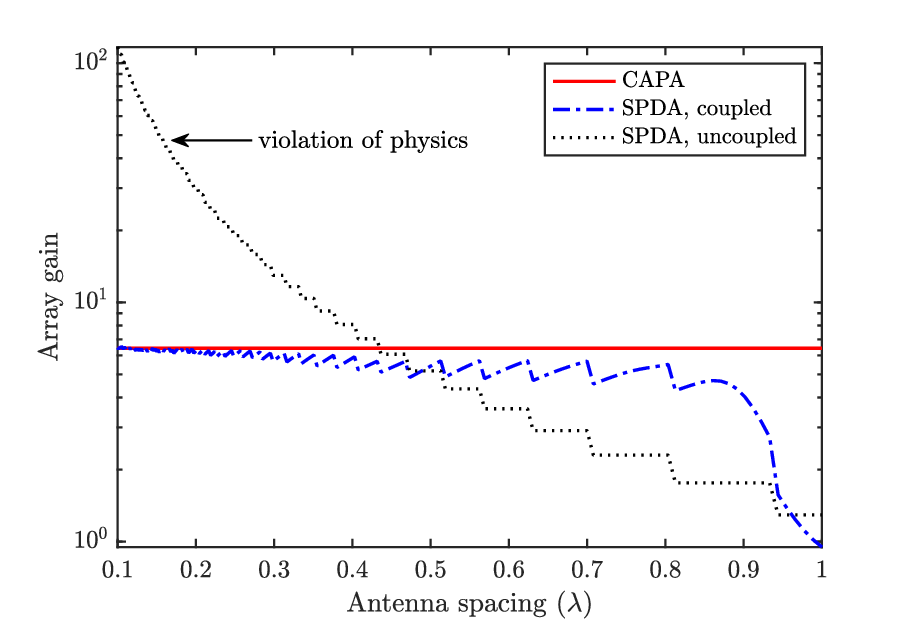}
    \caption{Array gain versus the antenna spacing of SPDAs.}
    \label{fig_array_gain_spacing}

    \centering
    \includegraphics[width=0.45\textwidth]{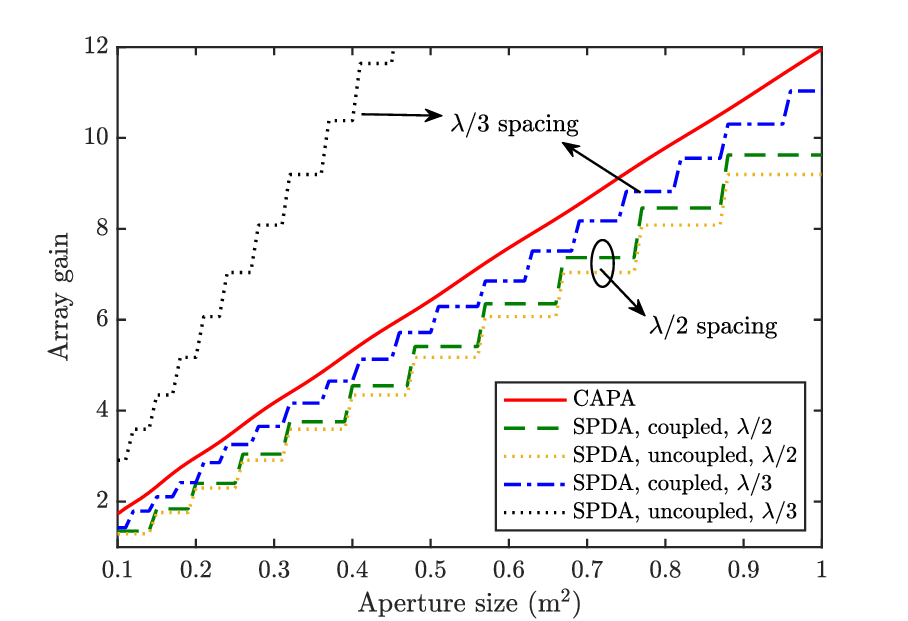}
    \caption{Array gain versus the array aperture.}
    \label{fig_array_gain_aperture}
\end{figure}

\subsection{Convergence of Proposed Methods}

We first evaluate the convergence performance of the proposed kernel approximation (KA) and conjugate gradient (CG) methods. First, the accuracy of both methods is highly dependent on the order of the Gauss-Legendre quadrature. In particular, the kernel approximation method exploits the Gauss-Legendre quadrature to approximate the radiation mutual coupling kernel, c.f. \eqref{proposed_approx_kernel}. Fig. \ref{fig_approx_accuracy} illustrates the true kernel and the approximated kernel under different Gauss-Legendre quadrature orders $M$ at 7.8 GHz. It can be observed that as $M$ increases, the approximated kernel gradually converges to the true kernel, validating the effectiveness of the proposed method. Furthermore, Fig. \ref{fig_GL_convergence} demonstrates the convergence of the array gain achieved by both methods with respect to $M$. Both methods eventually converge to the same value under different system parameters, which cross-validates the correctness of the results. We note that the CG method converges more smoothly. This may be because KA method exploits the \emph{model approximation}, i.e., approximating the coupling kernel itself, whereas the CG method applies \emph{numerical approximation}, i.e., approximating the final numerical implementation. Finally, Fig. \ref{fig_CG_convergence} shows the convergence of the CG method as iterations proceed. As the frequency increases, the CG method requires more iterations to converge. In contrast, the KA method is non-iterative, thus exhibiting significantly lower computational complexity. Given this advantage, the KA method is used to generate the numerical results in the following sections.

\begin{figure}[t]
    \centering
    \includegraphics[width=0.45\textwidth]{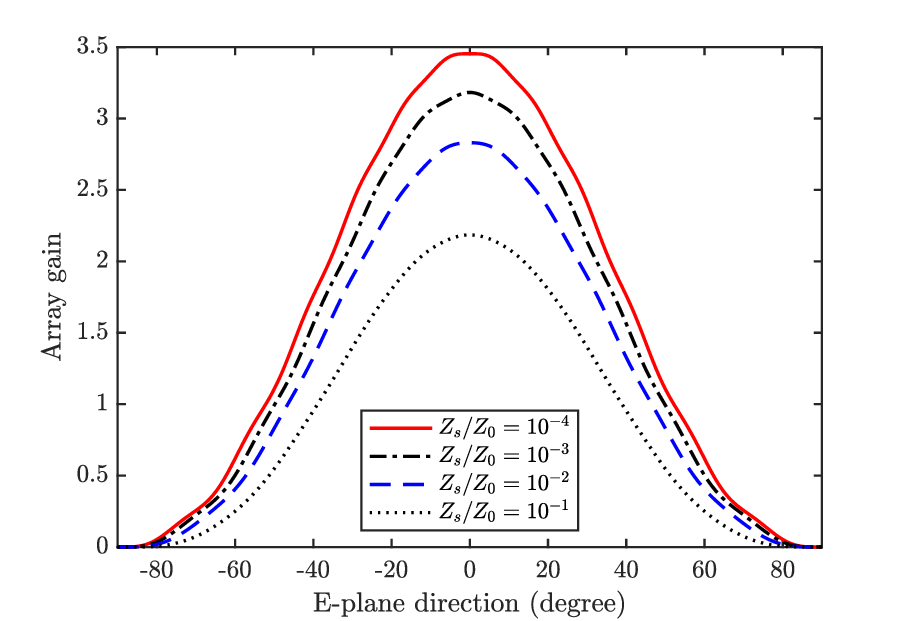}
    \caption{Array gain in the E-plane ($\theta = \pi/2$) with different surface resistance $Z_s$.}
    \label{fig_array_gain_E_plane}

    \centering
    \includegraphics[width=0.45\textwidth]{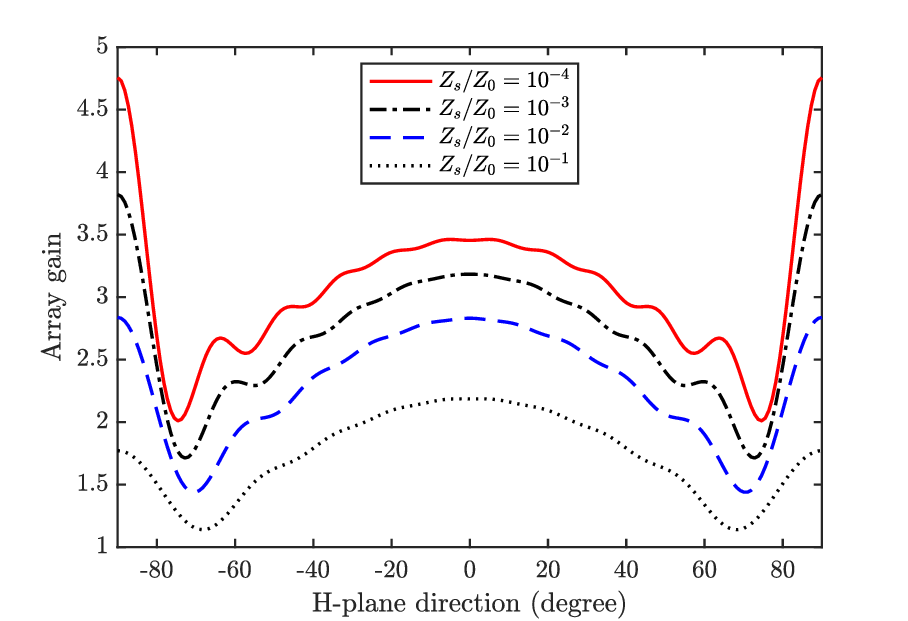}
    \caption{Array gain in the H-plane ($\theta = 0$) with different surface resistance $Z_s$.}
    \label{fig_array_gain_H_plane}
\end{figure}

\subsection{Array Gain}

We now evaluate the array gains under different setups, using both coupled and uncoupled SPDAs as benchmarks. The uncoupled benchmark is obtained by modifying the matrix $\mathbf{\Psi}$ defined in \eqref{SPDA_mutual_coupling_matrix}, keeping only its diagonal elements and setting all off-diagonal elements to zero.

Fig. \ref{fig_array_gain_spacing} depicts the optimal array gain versus the antenna spacing of SPDAs, where smaller spacing implies more antennas are deployed within a fixed aperture. As expected, when mutual coupling is considered, the SPDA array gain gradually converges to the CAPA performance as antenna spacing shrinks. This matches the physical expectation that CAPA is the ultimate limit of the SPDA. Nevertheless, the uncoupled model fails to capture the correct behavior. It underestimates performance at large spacing and significantly overestimates it at small spacing. Specifically, the array gain for the uncoupled SPDAs tends to increase unboundedly as spacing reduces, which is a violation of physics. Fig. \ref{fig_array_gain_aperture} further shows the array gain versus the aperture size $|\mathcal{S}|$. For coupled CAPAs, the array gain grows linearly as the aperture size increases. However, the SPDA array gain exhibits a stepwise increase, since a larger aperture does not always accommodate more antenna elements.

Fig. \ref{fig_array_gain_E_plane} and Fig. \ref{fig_array_gain_H_plane} investigate the array gains in the E-plane ($\theta = \pi/2$) and H-plane ($\theta = 0$), respectively, subject to different values of the surface resistance $Z_s$. In the E-plane, the array gain is maximized at front-fire ($\phi = 0$) and reduces to zero at end-fire ($\phi = \pm \pi/2$), where a larger surface resistance consistently leads to a lower array gain. The H-plane, however, exhibits a different behavior. As the direction approaches end-fire, the H-plane gain first decreases, as discussed in Section \ref{sec_array_gain}, but then reaches another peak at end-fire. This phenomenon is attributed to the windowing effect caused by the finite aperture. We also observe that as the surface resistance increases, these end-fire peaks become progressively lower than the front-fire peaks. Since a larger $Z_s$ essentially reduces the impact of mutual coupling, this observation implies that stronger mutual coupling can lead to a larger array gain at the end-fire, which is a form of superdirectivity.

\begin{figure}[t]
    \centering
    \includegraphics[width=0.45\textwidth]{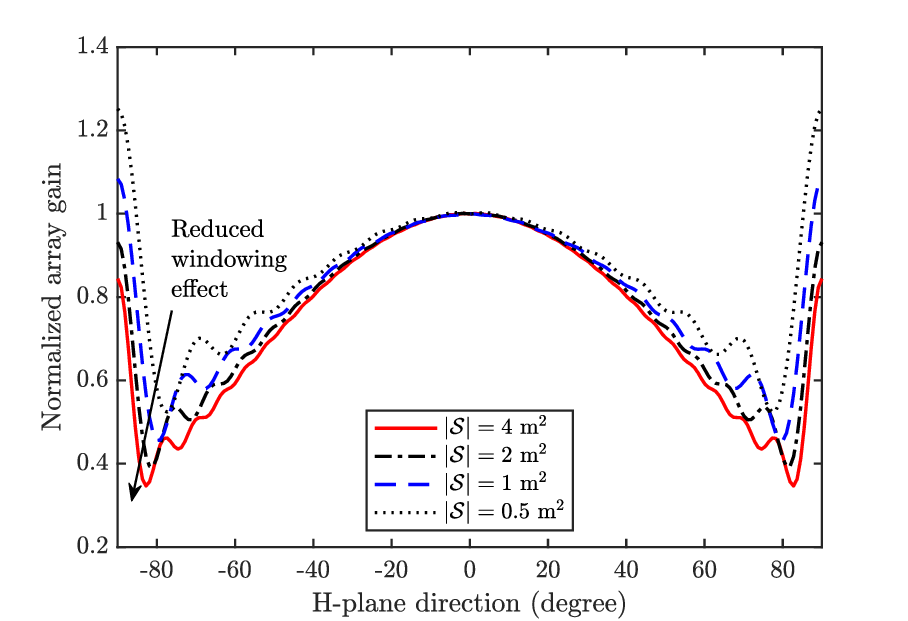}
    \caption{Normalized array gain in the H-plane ($\theta = 0$) with different aperture size.}
    \label{fig_array_gain_H_plane_2}
\end{figure}

Fig. \ref{fig_array_gain_H_plane_2} provides further insight into the H-plane gain under different aperture sizes. We observe that as the aperture size increases, the array gain at end-fire is reduced due to a less significant windowing effect. This result aligns with the analysis in Section \ref{sec_array_gain}, which indicates that for an infinitely large aperture, the array gain at end-fire becomes zero.

\subsection{Beampattern}

\begin{figure*}[t!]
    \centering
    \begin{subfigure}[t]{0.245\textwidth}
        \includegraphics[width=1\textwidth]{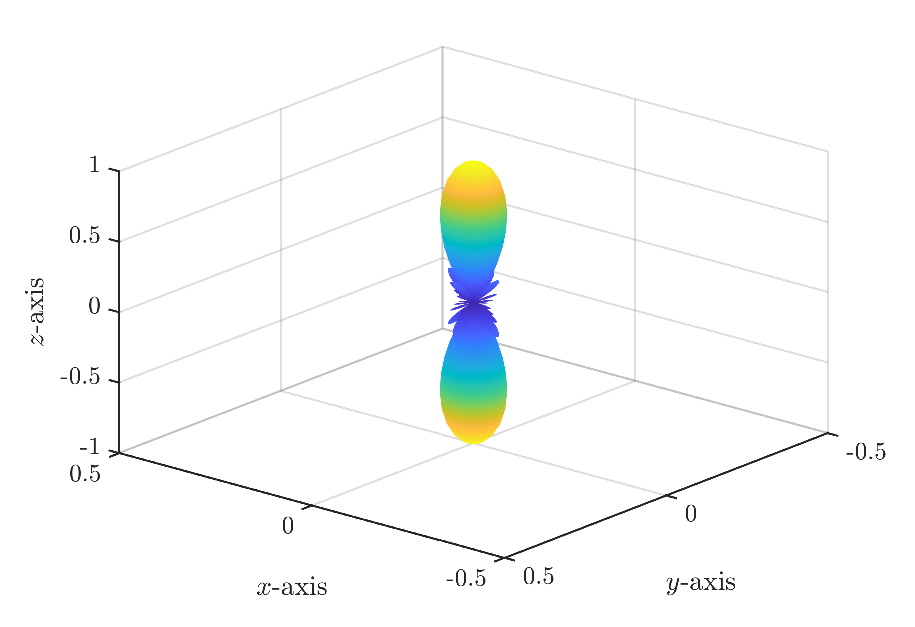}
        \caption{Front-fire beamforming, coupled.}
    \end{subfigure}
    \begin{subfigure}[t]{0.245\textwidth}
        \includegraphics[width=1\textwidth]{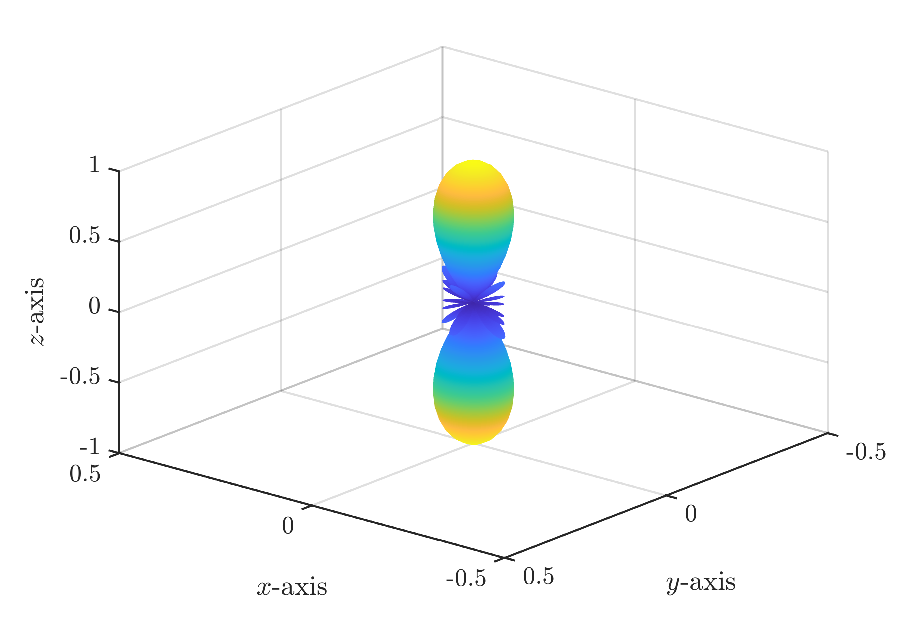}
        \caption{Front-fire beamforming, uncoupled.}
    \end{subfigure}
        \begin{subfigure}[t]{0.245\textwidth}
        \includegraphics[width=1\textwidth]{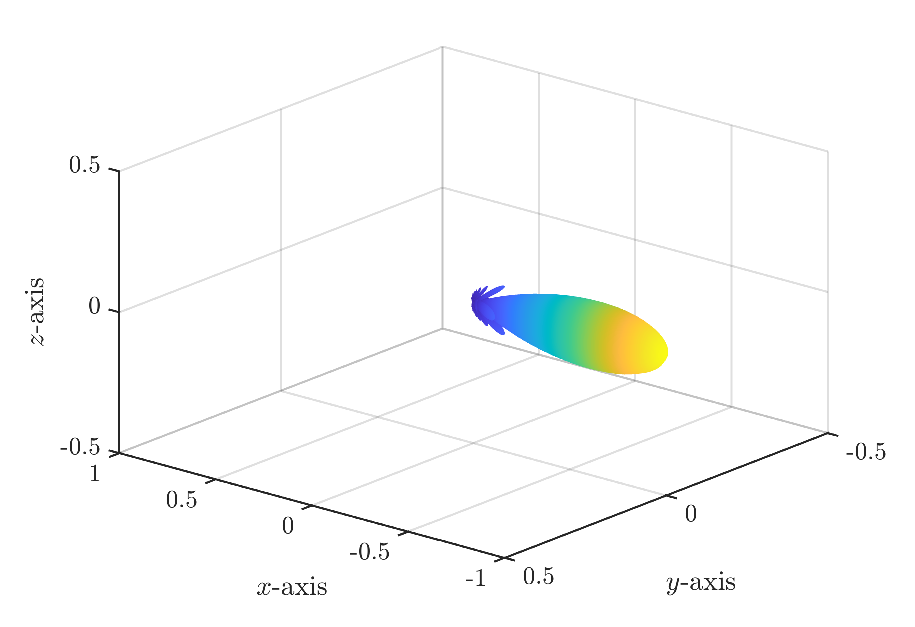}
        \caption{End-fire beamforming, coupled}
    \end{subfigure}
    \begin{subfigure}[t]{0.245\textwidth}
        \includegraphics[width=1\textwidth]{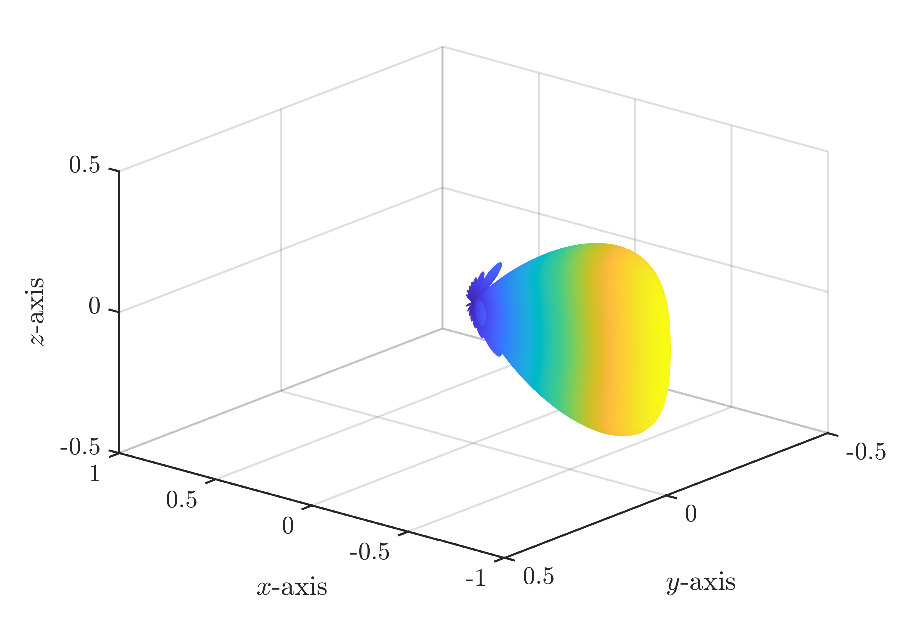}
        \caption{End-fire beamforming, uncoupled}
    \end{subfigure}
    \caption{Normalized beampatterns for coupled and uncoupled CAPAs. Front-fire beamforming is optimized for the $(\theta = 0, \phi = 0)$ direction, and end-fire beamforming is optimized for the $(\theta = 0, \phi = \pi/2)$ direction. Note that another end-fire direction $(\theta = \pi/2, \phi = \pi/2)$ exhibits zero gain due to polarization mismatch and is therefore not plotted.}
    \label{fig_beampattern}
    \vspace{-0.4cm}
\end{figure*}

Fig. \ref{fig_beampattern} illustrates the normalized beampatterns of coupled and uncoupled CAPAs, which align with the analysis in Section~\ref{sec_beampattern}. Compared to the uncoupled case, the coupled beampattern is filtered by the mutual coupling kernel. For both front-fire and end-fire beamforming, this mutual coupling effect leads to narrower beams. This narrowing effect is especially significant in the end-fire case, resulting in the realization of superdirectivity.

\section{Conclusions} \label{sec:conclusion}

This paper has developed a general physical model for mutual coupling in CAPAs, incorporating both polarization and surface dissipation loss. A primary finding is that polarization creates anisotropic coupling, meaning the coupling effect is direction-dependent. This discovery invalidates the conventional half-wavelength-spacing rule used in simpler models to eliminate coupling. Based on the proposed coupling model, the beamforming design was studied, where two methods were proposed to handle the difficulty of inverting the coupling kernel. The paper also extended the continuous mutual coupling model to SPDAs. Numerical results confirm that the performance of coupled SPDAs correctly converges to the CAPA limit as antenna spacing decreases. In contrast, uncoupled models are shown to violate physics, predicting unbounded array gain at small spacing. Furthermore, the analysis reveals that the coupled beampattern exhibits higher directivity than the uncoupled case, a phenomenon known as superdirectivity. 

This paper has primarily focused on uni-polarized and single-user CAPA systems. Although the underlying mutual coupling model is developed in a general tri-polarized form, the explicit beamforming design and numerical evaluation have been specialized to a fixed uni-polarized setting for analytical tractability. Several promising future research directions can be identified as follows:
\begin{itemize}
    \item \emph{Multi-Polarization Beamforming:} The paper develops the general tri-polarized coupling kernel but focuses the beamforming design on the uni-polarized case. Future work could extend the optimization framework to dual-polarized or tri-polarized systems to address the inter-polarization mutual coupling and fully exploit polarization diversity.
    
    \item \emph{Multi-User Design:} The proposed beamforming design targets a single receiver. A significant extension would be to address the multi-user scenario, developing beamforming solutions that manage inter-user interference in the presence of the complex coupling model.

    \item \emph{Antenna Current Profile Optimization:} The paper notes that the SPDA coupling matrix depends heavily on the current profile of each antenna element. This suggests a future co-design optimization problem, i.e, simultaneously optimizing the discrete beamforming weights and the physical current profile of the elements to manage or cancel mutual coupling.
    
    \item \emph{Full-Wave and Experimental Validation:} While this work develops a physics-based analytical model and validates its internal predictions numerically, an important next step is to benchmark the model against geometry-specific full-wave simulations and experimental measurements for finite practical arrays, including implementation-level effects such as feeding and matching constraints.

\end{itemize}

\section{Acknowledgment}
The authors would like to thank Dr. Chongjun Ouyang from Queen Mary University of London for the fruitful discussions regarding the initial modeling of mutual coupling.

\begin{appendices}

    \section{Radiation Power via Poynting Vector} \label{Poynting_vector_proof}

    From existing literature \cite{5446312, 9906802, zhang2023pattern}, the radial component of the Poynting vector is given by 
    \begin{equation}
        S(\mathbf{r}) = \frac{1}{2Z_0} \bm{e}_{\mathrm{rad}}^H(\mathbf{r}) \bm{e}_{\mathrm{rad}}(\mathbf{r}),
    \end{equation}
    where $\bm{e}_{\mathrm{rad}}(\mathbf{r})$ is defined in \eqref{radiation_field}. The radiation power can be calculated by integrating the Poynting vector over any closed surface that completely encloses the radiating aperture. To simplify the derivation, this surface is typically chosen at infinity, i.e., $R \triangleq \|\mathbf{r}\| \rightarrow \infty$, yielding the following radiation power:
    \begin{equation} \label{Poynting_vector_power}
        P_{\mathrm{rad}} = \lim_{R \rightarrow \infty} \oint_{\Omega} S(\mathbf{r}) R^2\, d \Omega,
    \end{equation}
    where $\Omega$ is the solid angle of $4\pi$ steradians.
    In this limit, the far-field approximation of the radiation field becomes exact, which is given by \cite{1386525}
    \begin{align} \label{far_field_E_field}
        \bm{e}_{\mathrm{rad}}(\mathbf{r}) = -j \kappa_0 Z_0 \frac{e^{j \kappa_0 R}}{4\pi R} \left( \mathbf{I}_3 - \frac{1}{\kappa_0^2} \mathbf{k} \mathbf{k}^T \right) \nonumber \\
        \times \underbrace{\int_{\mathcal{S}} \bm{j}_{\mathrm{t}}(\mathbf{s}) e^{-j \mathbf{k}^T \mathbf{s}} d \mathbf{s}}_{\triangleq \mathbf{J}_{\mathrm{t}}(\theta, \phi)},
    \end{align}
    where $\mathbf{k} = \kappa_0 \mathbf{d}(\theta, \phi)$. Plugging \eqref{far_field_E_field} into \eqref{Poynting_vector_power} and using the identity $\big(\mathbf{I}_3 - \frac{1}{\kappa_0^2} \mathbf{k} \mathbf{k}^T \big)^2 = \big(\mathbf{I}_3 - \frac{1}{\kappa_0^2} \mathbf{k} \mathbf{k}^T \big)$ gives
    \begin{align} \label{Poynting_vector_power_2}
        P_{\mathrm{rad}} & = \frac{\kappa_0^2 Z_0}{32 \pi^2} \lim_{R \rightarrow \infty} \oint_{\Omega} \left\| \left(\mathbf{I}_3 - \frac{1}{\kappa_0^2} \mathbf{k} \mathbf{k}^T \right) \mathbf{J}_{\mathrm{t}}(\theta, \phi) \right\|^2 \, d \Omega \nonumber \\
        & \hspace{-0.2cm} =  \frac{\kappa_0^2 Z_0}{32 \pi^2} \int_{\mathcal{S}} \int_{\mathcal{S}} \bm{j}_{\mathrm{t}}^H(\mathbf{s}) \left[ \oint_{\Omega} \left( \mathbf{I}_3 - \frac{1}{\kappa_0^2} \mathbf{k} \mathbf{k}^T \right) e^{j \mathbf{k}^T(\mathbf{s} - \mathbf{z})} d \Omega \right] \nonumber \\
        & \hspace{5cm} \times \bm{j}_{\mathrm{t}}(\mathbf{z}) d \mathbf{z} d \mathbf{s}.
    \end{align}
    Using the identity $\nabla \nabla e^{j \mathbf{k}^T \mathbf{s}} = -\mathbf{k} \mathbf{k}^T e^{j \mathbf{k}^T \mathbf{s}}$, the inner integral can be simplified into
    \begin{align} \label{o_integral}
        \oint_{\Omega} & \left( \mathbf{I}_3 - \frac{1}{\kappa_0^2} \mathbf{k} \mathbf{k}^T \right) e^{j \mathbf{k}^T \mathbf{s}} d \Omega \nonumber \\
        & = \left( \mathbf{I}_3 + \frac{1}{\kappa_0^2} \nabla \nabla \right) \oint_{\Omega} e^{j \mathbf{k}^T \mathbf{s}} d \Omega \nonumber \\
        & = \left( \mathbf{I}_3 + \frac{1}{\kappa_0^2} \nabla \nabla \right) \frac{4 \pi \sin(\kappa_0 \|\mathbf{s}\|)}{\kappa_0 \|\mathbf{s}\|} = \frac{16 \pi^2}{\kappa_0^2 Z_0} \Re \left\{ \mathbf{G}(\mathbf{s}) \right\}.
    \end{align}
    Substituting \eqref{o_integral} into \eqref{Poynting_vector_power_2} yields
    \begin{equation}
        P_{\mathrm{rad}} = \frac{1}{2} \int_{\mathcal{S}} \int_{\mathcal{S}}  \bm{j}_{\mathrm{t}}^H(\mathbf{s}) \Re \left\{\mathbf{G}(\mathbf{s} - \mathbf{z})\right\} \bm{j}_{\mathrm{t}}(\mathbf{z}) d \mathbf{z} d \mathbf{s},    
    \end{equation}
    which matches the radiation power in \eqref{tri_radiation_power}.

    \section{Proof of Proposition \ref{proposition_null}} \label{proposition_null_proof}

    To facilitate the derivation, we represent the coordinates by $s_x = r \sin \psi$ and $s_y = r \cos \psi$, where $r = \sqrt{s_x^2 + s_y^2} = \|\mathbf{s}\|$. The function $\varphi(\mathbf{s})$ can be rewritten as $\varphi(\mathbf{s}) = \sin(\kappa_0 r)/(4\pi r)$. Following the chain rule of derivative, the second-order derivative along the $y$-axis is given by 
    \begin{equation}
        \partial^2_y \varphi(\mathbf{s}) = \frac{d^2 \varphi}{d r^2} \cos^2 \psi + \frac{d \varphi}{d r} \frac{\sin^2 \psi}{r},
    \end{equation}  
    where the derivatives with respect to $r$ are given by 
    \begin{align}
        \frac{d \varphi}{d r} & = \frac{\kappa_0 r \cos(\kappa_0 r) - \sin(\kappa_0 r)}{4 \pi r^2}, \\
        \frac{d^2 \varphi}{d r^2} & = \frac{- \kappa_0^2 r^2 \sin(\kappa_0r) - 2 \kappa_0 r \cos(\kappa_0 r) + 2 \sin(\kappa_0 r)}{4 \pi r^3}.
    \end{align} 
    Based on the above results and defining $\epsilon = \kappa_0 r$, the null condition, i.e., $\varphi(\mathbf{s}) + \frac{1}{\kappa_0^2} \partial_y^2 \varphi(\mathbf{s}) = 0$ can be reformulated into the following compact form:
    \begin{equation}
        \sin^2 \psi \left( (\epsilon^2 - 3 ) \sin\epsilon + 3 \epsilon \cos \epsilon \right) + 2 (\sin \epsilon - \epsilon \cos \epsilon) = 0.
    \end{equation} 
    The proof is thus completed.
    
    \vspace{-0.3cm}
    \section{Proof of Theorem \ref{theorem_1}} \label{theorem_1_proof}

    This appendix provides the proof for the optimal continuous aperture beamformer stated in Theorem 1. The proof is derived using the calculus of variations based on the Karush-Kuhn-Tucker (KKT) conditions. The Lagrangian function of problem \eqref{problem_single_user} is given by 
    \begin{align}
        \mathcal{L}(w) = & \left| \int_{\mathcal{S}} h(\mathbf{s}) w(\mathbf{s}) d \mathbf{s} \right|^2 \nonumber \\
        & - \mu \left( \frac{1}{2} \int_{\mathcal{S}} \int_{\mathcal{S}} w(\mathbf{s}) c(\mathbf{s} - \mathbf{z}) w^*(\mathbf{z}) d \mathbf{z} d \mathbf{s} - P_{\mathrm{t}}\right),
    \end{align}
    where $\mu \ge 0$ is the Lagrange multiplier for the power constraint. 
    Following the principles of the calculus of variations, the optimal $w(\mathbf{s})$ maximizing $\mathcal{L}(w)$ is found where the first variation of $\mathcal{L}(w)$, denoted by $\delta \mathcal{L}(w, \delta w)$ is zero for any perturbation $\delta w$. The first variation is given by 
    \begin{align}
        \delta \mathcal{L}(w, \delta w) & = \left. \frac{d}{d \epsilon} \mathcal{L}(w + \epsilon \delta w) \right|_{\epsilon = 0}\!\!\! = 2\Re\left\{ \int_{\mathcal{S}} \delta w^*(\mathbf{s}) \chi(\mathbf{s}) d \mathbf{s} \right\},
    \end{align} 
    where
    \begin{align}
        \chi(\mathbf{s}) = \left(\int_{\mathcal{S}} h(\mathbf{z}) w(\mathbf{z}) d \mathbf{z}\right) h^*(\mathbf{s}) - \frac{\mu}{2} \int_{\mathcal{S}} c(\mathbf{s} - \mathbf{z}) w(\mathbf{z}) d \mathbf{z}.
    \end{align}
    For the functional $\mathcal{L}(w)$ to be at a maximum, we must have $\delta \mathcal{L}(w, \delta w) = 0$ for any arbitrary $\delta w$, implying $\chi(\mathbf{s})$ must be zero, i.e., 
    \begin{equation}
        \int_{\mathcal{S}} c(\mathbf{s} - \mathbf{z}) w(\mathbf{z}) d \mathbf{z} = \frac{2}{\mu}\left(\int_{\mathcal{S}} h(\mathbf{z}) w(\mathbf{z}) d \mathbf{z}\right) h^*(\mathbf{s}).
    \end{equation}

    Observing the above condition of the optimal solution, the optimal $w(\mathbf{s})$ must be a scaled version of $v(\mathbf{s})$ satisfying \eqref{Fredholm_equation}, i.e., $w_{\mathrm{opt}}(\mathbf{s}) = \alpha v(\mathbf{s})$. To find the scaling constant $\alpha$, we substitute this solution back into the power constraint \eqref{power_constraint}, which yields
    \begin{align}
        &\frac{\alpha^2}{2} \int_{\mathcal{S}} \int_{\mathcal{S}} v(\mathbf{s}) c(\mathbf{s} - \mathbf{z}) v^*(\mathbf{z}) d \mathbf{z} d \mathbf{s} = \frac{\alpha^2}{2} \int_{\mathcal{S}} h (\mathbf{z}) v(\mathbf{z}) d \mathbf{z} = P_{\mathrm{t}}.
    \end{align}    
    Therefore, we have $\alpha = \sqrt{\frac{2 P_{\mathrm{t}}}{\int_{\mathcal{S}} h (\mathbf{z}) v(\mathbf{z}) d \mathbf{z}}}$.
    This completes the proof.

    \section{Derivation of the Wavenumber-Domain \\ Coupling Kernel \eqref{WD_coupling_kernel}} \label{appendix_derivation_WD_coupling_kernel}

    This appendix derives the closed-form wavenumber-domain representation in \eqref{WD_coupling_kernel} for the radiation mutual-coupling kernel. Define $\Phi(\boldsymbol{\kappa}) \triangleq \mathcal{F}\{\varphi\}(\boldsymbol{\kappa})$. Then, from \eqref{radiation_mutual_coupling_kernel} and the Fourier-domain identity $\mathcal{F}\{\partial_y^2 \varphi\}(\boldsymbol{\kappa}) = -\kappa_y^2 \Phi(\boldsymbol{\kappa})$, the spectrum $C_{\mathrm{rad}}(\boldsymbol{\kappa})$ satisfies
    \begin{align}
    \label{eq:app_Crad_in_terms_of_Phi}
    C_{\mathrm{rad}}(\boldsymbol{\kappa})
    = & \ \kappa_0 Z_0 \left( \Phi(\boldsymbol{\kappa}) +\frac{1}{\kappa_0^2}\mathcal{F}\{\partial_y^2 \varphi\}(\boldsymbol{\kappa}) \right) \nonumber \\
    = & \ \kappa_0 Z_0 \left( 1-\frac{\kappa_y^2}{\kappa_0^2} \right) \Phi(\boldsymbol{\kappa}).
    \end{align}
    Therefore, it suffices to compute $\Phi(\boldsymbol{\kappa})$.

    Noting that $\varphi(\mathbf{s})=\Im\{g(\mathbf{s})\}$, we first invoke the Weyl identity at $s_z = 0$ to express $g(\mathbf{s})$ as \cite{chew1999waves}
    \begin{equation}
    g(\mathbf{s}) = \frac{j}{8 \pi^2} \iint_{-\infty}^{+\infty} \frac{e^{j \boldsymbol{\kappa}^T \mathbf{s}}}{\kappa_z(\boldsymbol{\kappa})}\,
    d\boldsymbol{\kappa},
    \quad
    \kappa_z(\boldsymbol{\kappa}) = \sqrt{\kappa_0^2-\|\boldsymbol{\kappa}\|^2}.
    \label{eq:app_weyl}
    \end{equation}
    For $\|\boldsymbol{\kappa}\|\le \kappa_0$, $\kappa_z(\boldsymbol{\kappa})$ is real. For $\|\boldsymbol{\kappa}\|>\kappa_0$, $\kappa_z(\boldsymbol{\kappa})=j\sqrt{\|\boldsymbol{\kappa}\|^2-\kappa_0^2}$ is purely imaginary. In the latter case, $j/\kappa_z(\boldsymbol{\kappa})$ is purely real and hence does not contribute to $\Im\{g(\mathbf{s})\}$. Consequently, $\varphi(\mathbf{s})$ is supported only on the disk $\|\boldsymbol{\kappa}\|\le \kappa_0$, which yields
    \begin{equation}
    \varphi(\mathbf{s}) = \frac{1}{(2\pi)^2}
    \iint_{\|\boldsymbol{\kappa}\|\le \kappa_0}
    \frac{e^{j \boldsymbol{\kappa}^T \mathbf{s}}}{2 \sqrt{\kappa_0^2-\|\boldsymbol{\kappa}\|^2}}\,
    d\boldsymbol{\kappa}.
    \label{eq:app_phi_weyl}
    \end{equation}
    Comparing \eqref{eq:app_phi_weyl} with the inverse Fourier transform gives
    \begin{equation}
    \Phi(\boldsymbol{\kappa}) =
    \begin{cases}
    \dfrac{1}{2\sqrt{\kappa_0^2-\|\boldsymbol{\kappa}\|^2}}, & \|\boldsymbol{\kappa}\|\le \kappa_0,\\[6pt]
    0, & \|\boldsymbol{\kappa}\|>\kappa_0.
    \end{cases}
    \label{eq:app_Phi_closed_form}
    \end{equation}
    Substituting \eqref{eq:app_Phi_closed_form} into \eqref{eq:app_Crad_in_terms_of_Phi} yields
    \begin{equation}
    C_{\mathrm{rad}}(\boldsymbol{\kappa})
    =
    \frac{Z_0\left(1-\kappa_y^2/\kappa_0^2\right)}
    {2\sqrt{1-\|\boldsymbol{\kappa}\|^2/\kappa_0^2}},
    \quad
    \|\boldsymbol{\kappa}\|\le \kappa_0,
    \label{eq:app_eq33_final}
    \end{equation}
    and $C_{\mathrm{rad}}(\boldsymbol{\kappa})=0$ for $\|\boldsymbol{\kappa}\|>\kappa_0$. This completes the derivation of \eqref{WD_coupling_kernel}.
    
    \vspace{-0.3cm}
        \section{Proof of Proposition \ref{proposition_infinite_array_gain}} \label{proposition_infinite_array_gain_proof}

    This proposition is proved using a wavenumber-domain analysis by transforming the optimization problem \eqref{problem_single_user} into the wavenumber domain. We begin by defining the 2D Fourier transform of the transmit beamformer $w(\mathbf{s})$ as $W(\boldsymbol{\kappa})$:
    \begin{equation} \label{windowed_FT}
        W(\boldsymbol{\kappa}) = \mathcal{F}\{w\}(\boldsymbol{\kappa}) = \iint_{-\infty}^{+\infty} w(\mathbf{s}) e^{-j \boldsymbol{\kappa}^T \mathbf{s}} d \mathbf{s}.
    \end{equation}  
    First, we reformulate the objective function when $|\mathcal{S}| \rightarrow +\infty$. Under the far-field assumption, the array gain $G$ can be expressed in terms of $W(\boldsymbol{\kappa})$ as
    \begin{align} \label{infinite_array_gain}
        \lim_{|\mathcal{S}| \rightarrow + \infty} G = & \frac{1}{P_{\mathrm{t}}}\left| \iint_{-\infty}^{+\infty} h(\mathbf{s}) w(\mathbf{s}) d \mathbf{s} \right|^2  \\
        = & \frac{1}{P_{\mathrm{t}}} \left| \iint_{-\infty}^{+\infty} \beta_{\mathrm{r}} w(\mathbf{s}) e^{-j \boldsymbol{\kappa}_{\mathrm{r}} \mathbf{s}} d \mathbf{s} \right|^2 = \frac{|\beta_{\mathrm{r}}|^2}{P_{\mathrm{t}}} \left|W(\boldsymbol{\kappa}_{\mathrm{r}})\right|^2. \nonumber
    \end{align}
    Next, we transform the power constraint \eqref{power_constraint}. This requires the Fourier transform of the overall coupling kernel $c(\mathbf{s})$, which is given by
    \begin{equation}
        C(\boldsymbol{\kappa}) = Z_s + C_{\mathrm{rad}}(\boldsymbol{\kappa}).
    \end{equation}
    By applying the Plancherel and convolution theorems, the transmit EM power $P_{\mathrm{em}}$ in the wavenumber domain is found to be
    \vspace{-0.2cm}
    \begin{equation}
        P_{\mathrm{em}} = \frac{1}{8 \pi^2} \iint_{-\infty}^{+\infty} C(\boldsymbol{\kappa}) \left|W(\boldsymbol{\kappa})\right|^2 d \boldsymbol{\kappa}
    \end{equation}
    By substituting these components, the original problem \eqref{problem_single_user} can be equivalently stated in the wavenumber domain as
    \begin{subequations}
        \begin{align}
            \max_{W(\boldsymbol{\kappa})} \quad & \left|W(\boldsymbol{\kappa}_{\mathrm{r}})\right|^2 \\
            \mathrm{s.t.} \quad & \frac{1}{8 \pi^2} \iint_{-\infty}^{+\infty} C(\boldsymbol{\kappa}) \left|W(\boldsymbol{\kappa})\right|^2 d \boldsymbol{\kappa} \le P_{\mathrm{t}}.
        \end{align}
    \end{subequations}
    The optimal $W(\boldsymbol{\kappa})$ can be obtained by Cauchy-Schwarz inequality as follows:
    \begin{align} \label{upper_bound_infinite}
        & \left|W(\boldsymbol{\kappa}_{\mathrm{r}})\right|^2 = \left| \frac{1}{\sqrt{C(\boldsymbol{\kappa}_{\mathrm{r}})}} \iint_{-\infty}^{+\infty} \sqrt{C(\boldsymbol{\kappa})} W(\boldsymbol{\kappa}) \delta(\boldsymbol{\kappa} - \boldsymbol{\kappa}_{\mathrm{r}}) d \boldsymbol{\kappa} \right|^2 \nonumber 
        \\
        & \le \frac{1}{C(\boldsymbol{\kappa}_{\mathrm{r}})} \underbrace{\iint_{-\infty}^{+\infty} C(\boldsymbol{\kappa}) \left| W(\boldsymbol{\kappa})\right|^2  d \boldsymbol{\kappa}}_{\le 8 \pi^2 P_{\mathrm{t}}} \underbrace{\iint_{-\infty}^{+\infty} \left|\delta(\boldsymbol{\kappa} - \boldsymbol{\kappa}_{\mathrm{r}})\right|^2  d \boldsymbol{\kappa}}_{= \delta(\mathbf{0})}  \nonumber \\
        & \le \frac{2 P_{\mathrm{t}} |\mathcal{S}| }{C(\boldsymbol{\kappa}_{\mathrm{r}})}.
    \end{align} 
    The last step exploit the identity $\delta(\mathbf{0}) = |\mathcal{S}| / (4 \pi^2)$, which is derived from the Fourier transform of a constant function.
    Plugging \eqref{upper_bound_infinite} into \eqref{infinite_array_gain} yields
    \begin{align}
        \lim_{|\mathcal{S}| \rightarrow + \infty} \!\! \frac{G_{\mathrm{opt}}}{|\mathcal{S}|} & = \frac{2|\beta_{\mathrm{r}}|^2}{Z_s + C_{\mathrm{rad}}(\boldsymbol{\kappa}_{\mathrm{r}})} \nonumber \\
        & \hspace{-0.8cm} = \frac{1}{4\pi^2} \left(\frac{\kappa_0}{R_0}\right)^2 \!\!\!\! \frac{Z_0^2(1 - \sin^2 \theta_0 \sin^2 \phi_0)^2 \cos \phi_0}{2 Z_s \cos \phi_0 + Z_0(1 - \sin^2\theta_0 \sin^2 \phi_0)},
    \end{align}
    which is derived based on \eqref{beta_factor} and \eqref{WD_coupling_kernel}.
    This completes the proof.

\end{appendices}

\balance
\bibliographystyle{IEEEtran}
\bibliography{reference/mybib}

@STRING{IEEE_J_IT         = "{IEEE} Trans. Inf. Theory"}

@STRING{IEEE_J_WCOML       = "{IEEE} Wireless Commun. Lett."}

@STRING{IEEE_J_JSAC       = "{IEEE} J. Sel. Areas Commun."}

@STRING{IEEE_J_COM        = "{IEEE} Trans. Commun."}

@STRING{IEEE_J_SP         = "{IEEE} Trans. Signal Process."}

@STRING{IEEE_J_WCOM       = "{IEEE} Trans. Wireless Commun."}

@STRING{IEEE_J_VT         = "{IEEE} Trans. Veh. Technol."}

@STRING{IEEE_WM_COM        = "{IEEE} Wireless Commun."}

@STRING{IEEE_Surveys        = "{IEEE} Commun. Surv. Tut."}

@STRING{IEEE_J_SP       = "{IEEE} Trans. Signal Process."}

@STRING{ICC        = "Proc. {IEEE} Int. Conf. Commun. ({ICC})"}

@STRING{IEEE_J_JSAIT       = "{IEEE} J. Sel. Areas Inf. Theory"}

@inproceedings{conference_version,
  author={Zhaolin Wang and Yuanwei Liu},
  title={Mutual Coupling Kernel Approximation for Continuous Aperture Beamforming},
  booktitle=ICC,
  address={Glasgow, United Kingdom},
  month=may,
  year={2026}
}

@ARTICLE{11122426,
  author={Wang, Zhaolin and Ouyang, Chongjun and Liu, Yuanwei},
  journal={IEEE Transactions on Wireless Communications}, 
  title={Beamforming Design for Continuous Aperture Array ({CAPA})-Based {MIMO} Systems}, 
  month=aug,
  year={2025, early access. doi: 10.1109/TWC.2025.3595157}
}

@article{wang2024tutorial,
  title={A tutorial on extremely large-scale MIMO for 6G: Fundamentals, signal processing, and applications},
  author={Wang, Zhe and Zhang, Jiayi and Du, Hongyang and Niyato, Dusit and Cui, Shuguang and Ai, Bo and Debbah, M{\'e}rouane and Letaief, Khaled B and Poor, H Vincent},
  journal=IEEE_Surveys,
  volume={26},
  number={3},
  pages={1560--1605},
  year={2024},
  month={3rd Quart.},
  publisher={IEEE}
}

@ARTICLE{10309946,
  author={Wang, Peng and Khormuji, Majid Nasiri and Popovic, Branislav M.},
  journal=IEEE_J_WCOM, 
  title={Beamforming Performances of Holographic Surfaces}, 
  year={2024},
  month=jun,
  volume={23},
  number={6},
  pages={5816-5831}
}

@article{yuan2023effects,
  title={Effects of mutual coupling on degree of freedom and antenna efficiency in holographic {MIMO} communications},
  author={Yuan, Shuai SA and others},
  author2={Yuan, Shuai SA and Chen, Xiaoming and Huang, Chongwen and Sha, Wei EI},
  journal={IEEE Open J. Antennas Propag.},
  volume={4},
  pages={237--244},
  year={2023},
  month=feb,
  publisher={IEEE}
}

@article{masouros2013large,
  title={Large-scale {MIMO} transmitters in fixed physical spaces: The effect of transmit correlation and mutual coupling},
  author={Masouros, Christos and Sellathurai, Mathini and Ratnarajah, Tharm},
  journal=IEEE_J_COM,
  volume={61},
  number={7},
  pages={2794--2804},
  year={2013},
  month=jul,
  publisher={IEEE}
}

@ARTICLE{4201034,
  author={Clerckx, Bruno and Craeye, Christophe and Vanhoenacker-Janvier, Danielle and Oestges, Claude},
  journal=IEEE_J_VT, 
  title={Impact of Antenna Coupling on 2 $\times$ 2 {MIMO} Communications}, 
  year={2007},
  volume={56},
  number={3},
  pages={1009-1018},
  month=may
}

@article{wong2014design,
  title={Design of unit cells and demonstration of methods for synthesizing Huygens metasurfaces},
  author={Wong, Joseph PS and Selvanayagam, Michael and Eleftheriades, George V},
  journal={Photonics Nanostructures-Fundam. Appl.},
  volume={12},
  number={4},
  pages={360--375},
  year={2014},
  publisher={Elsevier}
}

@INPROCEEDINGS{6404701,
  author={Müller, Ralf R. and Godana, Bruhtesfa E. and Sedaghat, Mohammad A. and Huber, Johannes B.},
  booktitle={Proc. IEEE Inf. Theory Workshop (ITW)}, 
  title={On channel capacity of communication via antenna arrays with receiver noise matching}, 
  year={2012},
  volume={},
  number={},
  pages={396-400}
}

@article{friedlander2020extended,
  title={The extended manifold for antenna arrays},
  author={Friedlander, Benjamin},
  journal=IEEE_J_SP,
  volume={68},
  pages={493--502},
  year={2020},
  month=jan,
  publisher={IEEE}
}

@inproceedings{yordanov2009arrays,
  title={Arrays of isotropic radiators-A field-theoretic justification},
  author={Yordanov, Hristomir and Ivrlac, Michel T and Russer, Peter and Nossek, Josef A},
  booktitle={Proc. ITG/IEEE Workshop on Smart Antennas},
  year={2009}
}

@book{chew1999waves,
  title={Waves and fields in inhomogenous media},
  author={Chew, Weng Cho},
  year={1999},
  publisher={John Wiley \& Sons}
}

@article{bjornson2024towards,
  title={Towards 6{G} {MIMO}: Massive spatial multiplexing, dense arrays, and interplay between electromagnetics and processing},
  author={Bj{\"o}rnson, Emil and Chae, Chan-Byoung and Heath Jr, Robert W and Marzetta, Thomas L and Mezghani, Amine and Sanguinetti, Luca and Rusek, Fredrik and Castellanos, Miguel R and Jun, Dongsoo and Demir, {\"O}zlem Tugfe},
  journal={arXiv preprint arXiv:2401.02844},
  year={2024}
}

@ARTICLE{10547020,
  author={D’Amico, Antonio Alberto and Sanguinetti, Luca},
  journal=IEEE_J_WCOM, 
  title={Holographic {MIMO} Communications: What is the Benefit of Closely Spaced Antennas?}, 
  year={2024},
  month=oct,
  volume={23},
  number={10},
  pages={13826-13840}
}

@ARTICLE{10158708,
  author={Akrout, Mohamed and Shyianov, Volodymyr and Bellili, Faouzi and Mezghani, Amine and Heath, Robert W.},
  journal=IEEE_J_JSAC, 
  title={Super-Wideband Massive {MIMO}}, 
  year={2023},
  month=aug,
  volume={41},
  number={8},
  pages={2414-2430}
}

@ARTICLE{1310320,
  author={Wallace, J.W. and Jensen, M.A.},
  journal=IEEE_WM_COM, 
  title={Mutual coupling in {MIMO} wireless systems: a rigorous network theory analysis}, 
  year={2004},
  month=jul,
  volume={3},
  number={4},
  pages={1317-1325}
}

@ARTICLE{5446312,
  author={Ivrlač, Michel T. and Nossek, Josef A.},
  journal={IEEE Trans. Circuits Syst. I: Regul. Pap.}, 
  title={Toward a Circuit Theory of Communication}, 
  year={2010},
  month=jul,
  volume={57},
  number={7},
  pages={1663-1683}
}

@article{zhu2024electromagnetic,
  title={Electromagnetic information theory: Fundamentals, modeling, applications, and open problems},
  author={Zhu, Jieao and Wan, Zhongzhichao and Dai, Linglong and Debbah, M{\'e}rouane and Poor, H Vincent},
  journal={IEEE Wireless Communications},
  volume={31},
  number={3},
  pages={156--162},
  year={2024},
  month=jun,
  publisher={IEEE}
}

@techreport{shewchuk1994introduction,
  title={An introduction to the conjugate gradient method without the agonizing pain},
  author={Shewchuk, Jonathan Richard},
  year={1994},
  institution={Carnegie-Mellon University}
}

@book{pozar2021microwave,
  title={Microwave Engineering},
  author={Pozar, David M},
  year={2012},
  publisher={Hoboken, NJ, USA: Wiley}
}

@ARTICLE{11006094,
  author={Pizzo, Andrea and Lozano, Angel},
  journal=IEEE_J_JSAIT, 
  title={Mutual Coupling in Holographic {MIMO}: Physical Modeling and Information-Theoretic Analysis}, 
  year={2025},
  month=may,
  volume={6},
  number={},
  pages={111-126}
}

@ARTICLE{9896943,
  author={Pizzo, Andrea and Lozano, Angel},
  journal=IEEE_J_WCOML, 
  title={On {Landau’s} Eigenvalue Theorem for Line-of-Sight {MIMO} Channels}, 
  year={2022},
  month=dec,
  volume={11},
  number={12},
  pages={2565-2569}
}

@article{liu2024capa,
  title={{CAPA}: Continuous-aperture arrays for revolutionizing {6G} wireless communications},
  author={Liu, Yuanwei and Ouyang, Chongjun and Wang, Zhaolin and Xu, Jiaqi and Mu, Xidong and Ding, Zhiguo},
  journal=IEEE_WM_COM,
  year={2025},
  month=aug,
  volume={32},
  number={4},
  pages={38-45}
}

@article{guo2024deep,
  title={Deep learning for beamforming in multi-user continuous aperture array ({CAPA}) systems},
  author={Guo, Jia and Liu, Yuanwei and Shin, Hyundong and Nallanathan, Arumugam},
  journal={arXiv preprint arXiv:2411.09104},
  year={2024}
}

@ARTICLE{10938678,
  author={Wang, Zhaolin and Ouyang, Chongjun and Liu, Yuanwei},
  journal=IEEE_J_COM, 
  title={Optimal Beamforming for Multi-User Continuous Aperture Array ({CAPA}) Systems}, 
  year={early access, Mar. 2025. doi: 10.1109/TCOMM.2025.3554644}
}

@book{olver2010nist,
	title        = {{NIST} Handbook of Mathematical Functions},
	author       = {Olver, Frank W and Lozier, Daniel W and Boisvert, Ronald F and Clark, Charles W},
	year         = 2010,
	publisher    = {Cambridge, U.K.: Cambridge Univ. Press}
}

@article{9906802,
	title        = {Wavenumber-Division Multiplexing in Line-of-Sight Holographic {MIMO} Communications},
	author       = {Sanguinetti, Luca and D’Amico, Antonio Alberto and Debbah, Merouane},
	year         = 2023,
	month        = apr,
	journal      = IEEE_J_WCOM,
	volume       = 22,
	number       = 4,
	pages        = {2186--2201}
}

@article{wan2023mutual,
	title        = {Mutual information for electromagnetic information theory based on random fields},
	author       = {Wan, Zhongzhichao and Zhu, Jieao and Zhang, Zijian and Dai, Linglong and Chae, Chan-Byoung},
	year         = 2023,
	month        = apr,
	journal      = IEEE_J_COM,
	publisher    = {IEEE},
	volume       = 71,
	number       = 4,
	pages        = {1982--1996}
}

@article{8585146,
	title        = {Horse (Electromagnetics) is More Important Than Horseman (Information) for Wireless Transmission},
	author       = {Migliore, Marco Donald},
	year         = 2019,
	month        = apr,
	journal      = {IEEE Trans. Antennas Propag.},
	volume       = 67,
	number       = 4,
	pages        = {2046--2055}
}

@article{jensen2008capacity,
	title        = {Capacity of the continuous-space electromagnetic channel},
	author       = {Jensen, Michael A and Wallace, Jon W},
	year         = 2008,
	month        = feb,
	journal      = {IEEE Trans. Antennas Propag.},
	publisher    = {IEEE},
	volume       = 56,
	number       = 2,
	pages        = {524--531}
}

@article{miller2000communicating,
	title        = {Communicating with waves between volumes: evaluating orthogonal spatial channels and limits on coupling strengths},
	author       = {Miller, David AB},
	year         = 2000,
	month        = apr.,
	journal      = {Appl. Opt.},
	publisher    = {Optica Publishing Group},
	volume       = 39,
	number       = 11,
	pages        = {1681--1699}
}

@article{bjornson2024enabling,
	author={Björnson, Emil and Kara, Ferdi and Kolomvakis, Nikolaos and Kosasih, Alva and Ramezani, Parisa and Salman, Murat Babek},
	journal={IEEE Open J. Commun. Soc.}, 
	title={Enabling 6G Performance in the Upper Mid-Band by Transitioning From Massive to Gigantic {MIMO}}, 
	year={2025},
	volume={6},
	number={},
	pages={5450-5463}
}

@article{zhang2023pattern,
	title        = {Pattern-division multiplexing for multi-user continuous-aperture {MIMO}},
	author       = {Zhang, Zijian and Dai, Linglong},
	year         = 2023,
	month        = aug,
	journal      = IEEE_J_JSAC,
	publisher    = {IEEE},
	volume       = 41,
	number       = 8,
	pages        = {2350--2366}
}

@article{1386525,
	title        = {Degrees of freedom in multiple-antenna channels: A signal space approach},
	author       = {Poon, A.S.Y. and Brodersen, R.W. and Tse, D.N.C.},
	year         = 2005,
	month        = feb,
	journal      = IEEE_J_IT,
	volume       = 51,
	number       = 2,
	pages        = {523--536}
}

@article{dardari2020communicating,
	title        = {Communicating with large intelligent surfaces: Fundamental limits and models},
	author       = {Dardari, Davide},
	year         = 2020,
	month        = nov,
	journal      = IEEE_J_JSAC,
	publisher    = {IEEE},
	volume       = 38,
	number       = 11,
	pages        = {2526--2537}
}

@article{10220205,
	title        = {Near-Field Communications: A Tutorial Review},
	author       = {Liu, Yuanwei and Wang, Zhaolin and Xu, Jiaqi and Ouyang, Chongjun and Mu, Xidong and Schober, Robert},
	year         = 2023,
	month        = aug,
	journal      = {IEEE Open J. Commun. Soc.},
	volume       = 4,
	number       = {},
	pages        = {1999--2049}
}

@article{10910020,
	title        = {Beamforming Optimization for Continuous Aperture Array ({CAPA})-based Communications},
	author       = {Wang, Zhaolin and Ouyang, Chongjun and Liu, Yuanwei},
	year={2025},
	month=jun,
	volume={24},
	number={6},
	pages={5099-5113},
	journal      = IEEE_J_WCOM
}

\end{document}